# Absence of spontaneous magnetic order of lattice spins coupled to itinerant interacting electrons in one and two dimensions

Daniel Loss[1], Fabio L. Pedrocchi[1], and Anthony J. Leggett[2]

[1]*Department of Physics, University of Basel, Klingelbergstrasse 82, CH-4056 Basel, Switzerland, and*
[2]*Department of Physics, University of Illinois at Urbana-Champaign,*
*1110 West Green Street, Urbana, Illinois 61801-3080, USA*

We extend the Mermin-Wagner theorem to a system of lattice spins which are spin-coupled to itinerant and interacting charge carriers. We use the Bogoliubov inequality to rigorously prove that neither (anti-) ferromagnetic nor helical long-range order is possible in one and two dimensions at any finite temperature. Our proof applies to a wide class of models including any form of electron-electron and single-electron interactions that are independent of spin. In the presence of Rashba or Dresselhaus spin-orbit interactions (SOI) magnetic order is not excluded and intimately connected to equilibrium spin currents. However, in the special case when Rashba and Dresselhaus SOIs are tuned to be equal, magnetic order is excluded again. This opens up a new possibility to control magnetism electrically.

Since the seminal work on phase transitions by Hohenberg [1] and Mermin and Wagner [2] it has become common knowledge that spontaneous order in low-dimensional systems is generically not possible at any finite temperature. In these studies, the use of the Bogoliubov inequality [3] was essential: Hohenberg used it to rule out superfluidity [1] and Mermin and Wagner to rule out magnetic order in Heisenberg spin systems [2] in dimensions $d < 3$. This approach is very powerful and was then applied to many different systems [4–10], including the Anderson and Kondo lattice models [11, 12].

For systems in the continuum, the weak coupling approximation is often applied leading to an effective exchange coupling between the localized spins which is of the RKKY-type [13]. RKKY interactions occur in many physical systems, prominent examples of present interest are heavy-fermion systems [14], diluted magnetic semiconductors [15–18], and nuclear spins in low-dimensional conducting nanostructures [19–21]. The latter system plays an important role as noise source for spin qubits in GaAs or InAs quantum dots [22–24], and much effort goes into understanding and controling the nuclear spin bath, with one possibility being to freeze out the nuclear noise by magnetic order [25, 26].

In contrast to the Heisenberg exchange, however, the RKKY interaction is long-ranged and thus is not covered by the original Mermin-Wagner theorem which requires the spin interactions to decay sufficiently fast with distance r (faster than $1/r^{2+d}$) [2]. Addressing precisely this issue, Bruno [10] was able to rule out in RKKY systems magnetic order in one dimension. A similar conclusion, however, for the two-dimensional counterpart appears still to be missing. Here we will fill this gap by rigorously proving the absence of order for a rather general class of systems which consist of lattice spins embedded in a continuum of itinerant electrons with which they interact by an isotropic on-site spin interaction. The allowed electron Hamiltonian $H_e$ is very general and may include electron-electron interactions as well as any single-particle potential (such as lattice or disorder potential) that does not depend on spin. For this class of models we prove then that in the thermodynamic limit ferro- and antiferromagnetic, as well as helical, long-range order of the lattice spins is excluded at any finite temperature in dimensions one and two. We show that this conclusion remains valid when short-range Heisenberg interaction between lattice spins is included. Our result also applies to the RKKY case, since this regime is obtained from the full one by lowest order perturbation expansion in the on-site spin interaction [13] including the full $H_e$ [26].

Moreover, we consider the effect of Rashba [27] and Dresselhaus [28] spin-orbit interactions (SOI) which explicitly break the spin symmetry. Our argument becomes then inconclusive and magnetic order cannot be excluded. While this finding is not unexpected it is remarkable that it is closely linked to the existence of equilibrium spin currents studied recently in spintronics [29–31]. Even more remarkably, we find that in the special case when Rashba ($\alpha$) and Dresselhaus ($\beta$) SOIs become equal, magnetic order is excluded again. Since $\alpha$ can be electrically tuned to $\beta$ [32–34], this opens up a new way to tune magnetism by electrical gates.

Finally, we note that the absence of spontaneous order proven here is valid only in the thermodynamic limit; thus, effective ordering in nanostructures of finite size at sufficiently low (but finite) temperatures is not in conflict with our findings.

*Model.* We consider a lattice $\{\mathbf{R}_j\}_{j=1}^{N_I}$ filled with $N_I$ spins $\hat{\mathbf{I}}_j = (\hat{I}_j^x, \hat{I}_j^y, \hat{I}_j^z)$ located at the sites $\mathbf{R}_j$. The lattice is embedded into a volume $\Omega$ containing $N_e$ itinerant electrons which couple to the lattice spins via on-site spin-spin interactions. The Hamiltonian for the entire system reads,

$$H = H_e + J\sum_{j=1}^{N_I} \hat{\mathbf{S}}_j \cdot \hat{\mathbf{I}}_j + h\sum_{j=1}^{N_I}(e^{-i\mathbf{Q}\cdot\mathbf{R}_j}\hat{I}_j^z + \text{h.c.}), \quad (1)$$

where $H_e = H_0 + V + U = \sum_{i=1}^{N_e} \hat{\mathbf{p}}_i^2/2m + \sum_{i<j}^{N_e} V_{ij} + \sum_{i=1}^{N_e} U(\hat{\mathbf{r}}_i)$ is the Hamiltonian describing the electron system. Here, $m$ is the mass and $\hat{\mathbf{p}}_i$ the momentum operator of the $i^{th}$ electron, $V_{ij} = V(\hat{\mathbf{r}}_i - \hat{\mathbf{r}}_j)$ the electron-electron interaction of electrons at positions $\hat{\mathbf{r}}_i$ and $\hat{\mathbf{r}}_j$, and $U(\hat{\mathbf{r}}_i)$ an arbitrary spin-independent single-electron potential. Typical examples for $U(\hat{\mathbf{r}}_i)$ are periodic lattice potentials, disorder potentials, electron-phonon interactions [35], etc. We remark that in contrast to previous work on lattice models [11, 12], we do not restrict the motion of the electrons to the sites of a lattice (tight binding limit) but allow them to move in the real space continuum. Further, $J$ denotes the coupling strength of the isotropic spin interaction at lattice site $\mathbf{R}_j$, $H_J = J\sum_{j=1}^{N_I}\hat{\mathbf{S}}_j \cdot \hat{\mathbf{I}}_j$, where $\hat{\mathbf{S}}_j \equiv \hat{\mathbf{S}}(\mathbf{R}_j)$ is the electron spin density operator $\hat{\mathbf{S}}(\mathbf{r}) = \sum_{i=1}^{N_e}\hat{\mathbf{s}}_i \delta(\mathbf{r} - \hat{\mathbf{r}}_i)$, with $\hat{\mathbf{s}}_i = (\hat{s}_i^x, \hat{s}_i^y, \hat{s}_i^z)$ being the spin-1/2 of the $i^{\text{th}}$ electron. The vector components of each spin, $\hat{s}_i^k$ and $\hat{I}_j^l$, satisfy standard spin commutation relations. Finally, to probe the order for the lattice spins $\hat{\mathbf{I}}_j$ we break the symmetry by an external (fictitious) field $h$ pointing in, say, the $z$ direction, which we let then go to zero at the end. This leads to an additional Zeeman term $H_Z(\mathbf{Q}) = h\sum_{j=1}^{N_I} e^{-i\mathbf{Q}\cdot\mathbf{R}_j}\hat{I}_j^z + \text{h.c.}$ To rule out ferromagnetic order we will choose $\mathbf{Q} = 0$, whereas to exclude antiferromagnetic order we will choose $\mathbf{Q}$ such that $e^{-i\mathbf{Q}\cdot\mathbf{R}} = +1$, if $\mathbf{R}$ connects sites from the same sublattice, and $e^{-i\mathbf{Q}\cdot\mathbf{R}} = -1$, if $\mathbf{R}$ connects sites from different sublattices.

To prove the absence of spontaneous order for the lattice spins $\hat{\mathbf{I}}_j$ we follow Ref. [2] and make use of the Bogoliubov inequality [3], which is an exact relation between two operators $A$, $C$, and a Hamiltonian $H$,

$$\frac{1}{2}\langle\{A, A^\dagger\}\rangle\langle[[C, H], C^\dagger]\rangle \geq k_B T|\langle[C, A]\rangle|^2. \quad (2)$$

Here, $\langle A \rangle = Tr e^{-H/k_B T} A / Tr e^{-H/k_B T}$ denotes the expectation value in a canonical ensemble, $T$ the temperature, $k_B$ the Boltzmann constant, and $\{A, B\} = AB + BA$ the anticommutator and $[A, B] = AB - BA$ the commutator. It is assumed that all expectation values are well-defined and exist in the thermodynamic limit defined by $N_e, N_I, \Omega \to \infty$ with electron density $n_e = N_e/\Omega$ and density of lattice spins $n_I = N_I/\Omega$ finite.

*Proof*- The strategy of the proof consists of using the Bogoliubov inequality to derive an upper bound for the order parameter corresponding to the phase transition we want to discuss. If this bound turns out to be in contradiction with the presence of long-range magnetic order, then the absence of the corresponding phase transition is rigorously demonstrated. The success of the procedure depends crucially on the choice of the operators $A$ and $C$ in (2). As we shall see, the appropriate choice for our case is given by

$$C_\mathbf{q} = \hat{S}_{-\mathbf{q}}^- + \hat{I}_{-\mathbf{q}}^- + \hat{S}_{-\mathbf{q}}^+ + \hat{I}_{-\mathbf{q}}^+, \quad A_\mathbf{q} = \hat{I}_{\mathbf{q}+\mathbf{Q}}^+ + \hat{I}_{\mathbf{q}-\mathbf{Q}}^+, \quad (3)$$

where the Fourier transforms are given by $\hat{\mathbf{S}}_\mathbf{q} = \sum_{i=1}^{N_e} e^{-i\mathbf{q}\cdot\hat{\mathbf{r}}_i}\hat{\mathbf{s}}_i$ and $\hat{\mathbf{I}}_\mathbf{q} = \sum_{j=1}^{N_I} e^{-i\mathbf{q}\cdot\mathbf{R}_j}\hat{\mathbf{I}}_j$ [36], and where $B^\pm \equiv B^x \pm iB^y$. Note that $C_\mathbf{q}$ and $A_\mathbf{q}$ are not Hermitian in general. Since the Bogoliubov inequality (2) is valid for any wave vector $\mathbf{q}$, it can be generalized to

$$\frac{1}{2}\sum_\mathbf{q}\langle\{A_\mathbf{q}, A_\mathbf{q}^\dagger\}\rangle \geq k_B T \sum_\mathbf{q} \frac{|\langle[C_\mathbf{q}, A_\mathbf{q}]\rangle|^2}{\langle[[C_\mathbf{q}, H], C_\mathbf{q}^\dagger]\rangle}, \quad (4)$$

where the sum runs over all $\mathbf{q}$'s in the first Brillouin zone of the reciprocal lattice. We note that the above choice for $C_\mathbf{q}$ and $A_\mathbf{q}$ is essential also for the following reason. Besides the fact that $\sum_\mathbf{q}\langle[C_\mathbf{q}, A_\mathbf{q}]\rangle$ can be expressed in terms of the lattice spin magnetization, the generally complicated interaction terms $V$ and $U$ in $H_e$ simply drop out of the calculation since they commute with $C_\mathbf{q}$,

$$[C_\mathbf{q}, H_e] = [\hat{S}_{-\mathbf{q}}^- + \hat{S}_{-\mathbf{q}}^+, H_0]. \quad (5)$$

This simplification is a crucial advantage of first over second quantization formalism since spin and position operators of the electrons trivially commute. (Note, however, that the expectation values still contain the full Hamiltonian including $U$ and $V$.) Hence, our proof goes through for any form of the potentials $V$ and $U$ as long as they are spin independent.

We now focus on the various terms in Eq. (4) and find bounds for them. Here, we outline only the main steps of the calculations and defer details to the Appendix [37]. As a first step, let us evaluate the double commutator on the right-hand side of inequality (4). By virtue of the commutation relation $[\hat{S}_{-\mathbf{q}}^\pm, H_0] = -\frac{\mathbf{q}}{2m}\sum_i \hat{s}_i^\pm\{\hat{\mathbf{p}}_i, e^{i\mathbf{q}\cdot\hat{\mathbf{r}}_i}\}$, we obtain that $[[C_\mathbf{q}, H_e], C_\mathbf{q}^\dagger] = \frac{1}{m}N_e q^2$. The part of the double commutator with $H_J$ vanishes since $[C_\mathbf{q}, H_J] = 0$. Indeed, $[\hat{\mathbf{S}}_{-\mathbf{q}}^\pm, H_J] = i\sum_{i,j} e^{i\mathbf{q}\cdot\hat{\mathbf{r}}_i}\delta(\hat{\mathbf{r}}_i - \mathbf{R}_j)(\hat{\mathbf{I}}_j \times \hat{\mathbf{s}}_i)^\pm$, and thus $[\hat{\mathbf{S}}_{-\mathbf{q}}^\pm, H_J] = -[\hat{\mathbf{I}}_{-\mathbf{q}}^\pm, H_J]$. After some calculations (see [37]) we find that $[[C_\mathbf{q}, H_z(\mathbf{Q})], C_\mathbf{q}^\dagger] = -4h(\sum_j e^{-i\mathbf{Q}\cdot\mathbf{R}_j}\hat{I}_j^z + \text{h.c.})$. Hence,

$$\langle[[C_\mathbf{q}, H], C_\mathbf{q}^\dagger]\rangle = N_e\left(\frac{1}{m}q^2 - 4h\frac{N_I}{N_e}m_I^z(\mathbf{Q})\right), \quad (6)$$

where the lattice spin magnetization appearing in Eq. (6), which we identify as the order parameter, is defined by $m_I^z(\mathbf{Q}) = \frac{1}{N_I}\langle\sum_j e^{-i\mathbf{Q}\cdot\mathbf{R}_j}\hat{I}_j^z + e^{i\mathbf{Q}\cdot\mathbf{R}_j}\hat{I}_j^z\rangle$. The commutator on the right-hand side of inequality (4) can also be expressed in terms of $m_I^z(\mathbf{Q})$,

$$\langle[C_\mathbf{q}, A_\mathbf{q}]\rangle = -2N_I m_I^z(\mathbf{Q}). \quad (7)$$

Finally, the sum on the left-hand side of Eq. (4) can be bounded as follows,

$$\sum_{\mathbf{q}} \langle \{A_{\mathbf{q}}, A_{\mathbf{q}}^{\dagger}\}\rangle = 2N_I \sum_j \langle\{\hat{I}_j^+, \hat{I}_j^-\}(1+\cos(\mathbf{Q}\cdot\mathbf{R}_j))\rangle$$
$$\leq 4N_I^2(2I)^2, \quad (8)$$

where we have used that $\sum_{\mathbf{q}} e^{i\mathbf{q}\cdot(\mathbf{R}_i-\mathbf{R}_j)} = N_I \delta_{\mathbf{R}_i,\mathbf{R}_j}$, and $\langle\{\hat{I}_j^+,\hat{I}_j^-\}\rangle \leq (2I)^2$. Using Eqs. (6), (7), and (8), we obtain from the Bogoliubov inequality (4)

$$4N_I^2(2I)^2/2 \geq k_B T \sum_{\mathbf{q}} \frac{4N_I^2 m_I^z(\mathbf{Q})^2}{\langle[[C_{\mathbf{q}},H],C_{\mathbf{q}}^{\dagger}]\rangle}. \quad (9)$$

Our goal is to rule out spontaneous magnetization in the lattice spin system, therefore we are interested in the behavior of the order parameter $m_I^z(\mathbf{Q})$ in the limit of vanishing external field, i.e., $h \to 0$, *after* we have taken the thermodynamic limit. We need to distinguish two cases: i) $m_I^z(\mathbf{Q}) = 0$, $\forall h$ around $h = 0$; ii) $m_I^z(\mathbf{Q}) \neq 0$, $\forall h$ around $h = 0$. If i) is satisfied, there is no order and the proof is completed. If ii) is satisfied, we need to show that $\lim_{h\to 0} m_I^z(\mathbf{Q}) = 0$ follows from inequality (9) in the thermodynamic limit. In this limit, the sum can be replaced by an integral,

$$(2I)^2 \geq \frac{k_B T N_I v}{N_e(2\pi)^d} \int\limits_{|\mathbf{q}|\leq|\mathbf{q}_c|} d^d q \, \frac{m_I^z(\mathbf{Q})^2}{\frac{q^2}{2m}+|\nu h m_I^z(\mathbf{Q})|}, \quad (10)$$

where $\nu = 2N_I/N_e$, $\mathbf{q}_c$ is an arbitrary cut-off vector lying in the first Brillouin zone, $v = \Omega/N_I$, and we have used that $\langle[[C_{\mathbf{q}},H],C_{\mathbf{q}}^{\dagger}]\rangle \leq N_e(q^2/m + |2\nu h m_I^z(\mathbf{Q})|)$. In the one-dimensional case ($d = 1$), Eq. (10) gives

$$\frac{\lambda_1 \sqrt{|h|}}{T}\left[\arctan\left(\frac{|\mathbf{q}_c|}{\sqrt{2m|\nu h m_I^z(\mathbf{Q})|}}\right)\right]^{-1} \geq \frac{m_I^z(\mathbf{Q})^2}{\sqrt{|m_I^z(\mathbf{Q})|}}, \quad (11)$$

where $\lambda_1 = \pi(2I)^2 n_e \sqrt{\nu}/(k_B \sqrt{2m})$. In the limit $h \to 0$, the left-hand side of inequality (11) vanishes and this implies that $\lim_{h\to 0} m_I^z(\mathbf{Q}) = 0$. The two-dimensional case can be treated in a similar way. For $d = 2$, inequality (10) leads to the following relation

$$\frac{\lambda_2}{T}\left[\log\left(1+\frac{|\mathbf{q}_c|^2}{2m|\nu h m_I^z(\mathbf{Q})|}\right)\right]^{-1} \geq m_I^z(\mathbf{Q})^2, \quad (12)$$

where $\lambda_2 = 2\sqrt{2}\lambda_1/\sqrt{\nu m}$. It follows from inequality (12) that $\lim_{h\to 0} m_I^z(\mathbf{Q}) = 0$ here, too. Since our arguments were independent of the choice of $\mathbf{Q}$, we have proven that neither ferromagnetic nor antiferromagnetic long-range order of the lattice spins is possible at any finite temperature $T > 0$ in one and two dimensions.

The absence of order can be traced back to the increased fluctuations in the lattice spin system in lower dimensions. These fluctuations, in turn, have their origin in the kinetic energy of the electrons, as one can explicitly see from Eq. (10) where the term $q^2/2m$ is responsible for the divergency in above q-integrals for $d = 1$ and 2.

Next, we show that helical long-range order of the lattice spins is also excluded. The strategy of the proof remains the same and we shall be brief (for details see [37]). To study this type of order, we consider the symmetry breaking Zeeman term $\widetilde{H}_Z(\mathbf{Q}) = \sqrt{2/3}h\sum_j e^{-i\mathbf{Q}\cdot\mathbf{R}_j}\hat{I}_j^+ + h.c.$ and the magnetic order parameter $m_I^{\perp}(\mathbf{Q}) = \sqrt{2/3}\frac{1}{N_I}\langle\sum_j e^{-i\mathbf{Q}\cdot\mathbf{R}_j}\hat{I}_j^+ + h.c.\rangle$ which corresponds to a spin helix in the $xy$-plane. Note that the spin part of Hamiltonian (1) is isotropic and consequently all choices for the helix are equivalent. The operators $\widetilde{C}_{\mathbf{q}}$ and $\widetilde{A}_{\mathbf{q}}$ for the Bogoliubov inequality (4) are now chosen to be

$$\widetilde{C}_{\mathbf{q}} = \hat{S}_{-\mathbf{q}}^z + \hat{I}_{-\mathbf{q}}^z \quad \text{and} \quad \widetilde{A}_{\mathbf{q}} = \frac{1}{\sqrt{3}}\left(\hat{I}_{\mathbf{q}+\mathbf{Q}}^+ - \hat{I}_{\mathbf{q}-\mathbf{Q}}^-\right). \quad (13)$$

The double commutator on the right-hand side of Eq. (4) becomes then $\langle[[\widetilde{C}_{\mathbf{q}},H],\widetilde{C}_{\mathbf{q}}^{\dagger}]\rangle = N_e(q^2/4m - \nu h m_I(\mathbf{Q})/2)$. Since $\langle[\widetilde{C}_{\mathbf{q}},\widetilde{A}_{\mathbf{q}}]\rangle = (N_I/\sqrt{2})m_I^{\perp}(\mathbf{Q})$ and $\sum_{\mathbf{q}}\langle\{\widetilde{A}_{\mathbf{q}},\widetilde{A}_{\mathbf{q}}^{\dagger}\}\rangle \leq 2N_I^2(2I)^2$, Eq. (4) takes in the thermodynamic limit exactly the same form as Eq. (10), where $m_I^z(\mathbf{Q})$ must be replaced by $m_I^{\perp}(\mathbf{Q})$. We thus conclude that $\lim_{h\to 0} m_I^{\perp}(\mathbf{Q}) = 0$ for any $\mathbf{Q}$ and hence long-range helical order is also excluded in one and two dimensions at any $T > 0$ [38].

As a further generalization, short-range impurity-spin Heisenberg interaction $H_{\mathcal{I}} = \sum_{i,j} \mathcal{I}_{ij}\hat{\mathbf{I}}_i \cdot \hat{\mathbf{I}}_j$ is added to Hamiltonian (1). When the couplings $\mathcal{I}_{ij}$ satisfy $1/N_I \sum_{ij}|\mathcal{I}_{ij}|(\mathbf{R}_i - \mathbf{R}_j)^2 < \infty$, then both proofs to exclude (anti-) ferromagnetic and helical ordering remain valid and lead to Eq. (10) with renormalized mass $m^* = m/(1 + 8mI^2\frac{n_I}{n_e}\frac{1}{N_I}\sum_{ij}|\mathcal{I}_{ij}|(\mathbf{R}_i - \mathbf{R}_j)^2)$ [37].

*Presence of spin-orbit interaction.* Next we investigate the question of magnetic order in a low-dimensional electron gas in the presence of Rashba [27] and/or Dresselhaus [28] spin-orbit interaction which break the rotational spin symmetry of the Hamiltonian (1) explicitly. The spin-orbit Hamiltonian is given by $H_{\text{SO}} = H_R + H_D$, with $H_R = \alpha \sum_{i=1}^{N_e}(\hat{p}_i^y \hat{s}_i^x - \hat{p}_i^x \hat{s}_i^y)$, $H_D = \beta \sum_{i=1}^{N_e}(\hat{p}_i^x \hat{s}_i^x - \hat{p}_i^y \hat{s}_i^y)$, where $\alpha$ ($\beta$) is the Rashba (Dresselhaus) coefficient. Using Eq. (3) for $C_{\mathbf{q}}$, we obtain $[[C_{\mathbf{q}},H_{\text{SO}}],C_{\mathbf{q}}^{\dagger}] = 4m\alpha\hat{j}_{\mathbf{q}=\mathbf{0},x}^y + 4m\beta\hat{j}_{\mathbf{q}=\mathbf{0},y}^y$, where we have defined the spin-current density operator as $\hat{\mathbf{j}}^{\alpha}(\mathbf{r}) = \frac{1}{2m}\sum_{i=1}^{N_e}\hat{s}_i^{\alpha}\{\hat{\mathbf{p}}_i, \delta(\hat{\mathbf{r}}_i - \mathbf{r})\}$ and its corresponding Fourier component $\hat{\mathbf{j}}_{\mathbf{q}}^{\alpha} = \frac{1}{2m}\sum_i \hat{s}_i^{\alpha}\{\hat{\mathbf{p}}_i, e^{-i\mathbf{q}\cdot\hat{\mathbf{r}}_i}\}$. These spin currents may lead to an intrinsic cut-off for the fluctuations in $q$, and thus help to establish order. To see this, we evaluate now the spin currents perturbatively around the free electron limit, i.e. $U,V,J = 0$, and at $T = 0$ [39],

$$\langle j_{\mathbf{q}=\mathbf{0},x}^y\rangle_0 = \Omega\frac{mE_F}{4\pi}\alpha \quad (14)$$

$$\langle j_{\mathbf{q}=\mathbf{0},y}^y\rangle_0 = -\Omega\frac{mE_F}{4\pi}\beta, \quad (15)$$

where $E_F$ is the Fermi energy and the results are valid in the regime $m\alpha^2, m\beta^2 \ll \hbar^2 E_F$ [40]. Performing now a perturbative expansion in the parameters $V, U, J, T$ around above free case, we conclude that $\langle \hat{j}_x^y \rangle \neq 0$ and $\langle \hat{j}_x^y \rangle \neq 0$ [41]. (In passing we note that in the stationary and homogeneous limit, the spin-currents satisfy the relations $\langle \hat{j}_x^x \rangle = -\langle \hat{j}_y^y \rangle$ and $\langle \hat{j}_x^y \rangle = -\langle \hat{j}_y^x \rangle$ due to a generalized continuity equation; see [37].) As a consequence, the commutator $\langle [[C_{\mathbf{q}}, H_{\text{SO}}], C_{\mathbf{q}}^\dagger] \rangle$ appearing in Eq. (9) does not vanish anymore and thus provides an intrinsic cut-off to the q-integral [cf. Equation (10)]. Hence, the bound for the order parameter we extract from inequality (10) is a constant which does not vanish in the limit $h \to 0$. Thus, our argument becomes inconclusive and we cannot rule out (anti-) ferromagnetic order in this case.

Similarly, for helical order our argument remains inconclusive, since $[[\widetilde{C}_{\mathbf{q}}, H_{SO}], \widetilde{C}_{\mathbf{q}}^\dagger] = m\alpha(\hat{j}_{\mathbf{q}=0,x}^y - \hat{j}_{\mathbf{q}=0,y}^x) + m\beta(\hat{j}_{\mathbf{q}=0,y}^y - \hat{j}_{\mathbf{q}=0,x}^x)$, which, will not vanish in general.

Next, let us consider the special case $\alpha = \beta$ where new symmetries emerge [42]. Then, the leading terms, Eqs. (14), (15), cancel, indicating that the physics changes dramatically. Indeed, by making use of the "gauge transformation" $U = e^{i \sum_k \hat{\mathbf{A}}_k \cdot \hat{\mathbf{r}}_k}$, where $\hat{\mathbf{A}}_k = -\alpha m(\hat{s}_k^x - \hat{s}_k^y)(1,1,0)$, to remove the SOI from the Hamiltonian, we can prove as before [37] that (anti-) ferromagnetic order in the $z$ direction can now be excluded rigorously for any $T > 0$ and $d = 1, 2$. Similarly, we can rule out helical ordering described by the order parameter $m_I^{\perp'} = \frac{1}{N_I} \langle \sum_j e^{-i\mathbf{Q}\cdot\mathbf{R}_j} \hat{I}_j^{+'} + h.c. \rangle$ with $\mathbf{Q} = \sqrt{2}\alpha m(1,1,0)$ [for rotated coordinates $(x, y, z) \to (x', y', z') = (z, (x+y)/\sqrt{2}, (x-y)/\sqrt{2})$; see [37]].

Thus, quite remarkably, this spin-orbit effect suggests the control of magnetism by electrical gates, namely, by tuning the Rashba SOI ($\alpha$) [32–34] from the regime $\alpha \neq \beta$ (ordering not excluded) to $\alpha = \beta$ (ordering excluded).

*Conclusions.* We proved an extension of the Mermin-Wagner theorem for lattice spins interacting with itinerant electrons, and showed that spontaneous order of the lattice spins is ruled out in one and two dimensions at finite temperature. In the presence of Rashba ($\alpha$) and Dresselhaus ($\beta$) spin-orbit interactions, however, spontaneous order could not be excluded, unless for $\alpha = \beta$.

We would like to thank K. van Hoogdalem, D. Rainis, and D. Stepanenko for useful discussions. We acknowledge support from the Swiss NF, NCCRs Nanoscience and QSIT, SOLID, and DARPA.

# Supplementary Material to "Absence of spontaneous magnetic order of lattice spins coupled to itinerant interacting electrons in one and two dimensions"

Daniel Loss[1], Fabio L. Pedrocchi[1], and Anthony J. Leggett[2]

[1]*Department of Physics, University of Basel, Klingelbergstrasse 82, CH-4056 Basel, Switzerland, and*
[2]*Department of Physics, University of Illinois at Urbana-Champaign,*
*1110 West Green Street, Urbana, Illinois 61801-3080, USA*

## I. MODEL HAMILTONIAN

To make the Appendix largely self-contained we restate the problem defined in the main text briefly. We consider a lattice $\{\mathbf{R}_j\}_{j=1}^{N_I}$ filled with $N_I$ spins $\hat{\mathbf{I}}_j = (\hat{I}_j^x, \hat{I}_j^y, \hat{I}_j^z)$ located at the sites $\mathbf{R}_j$. The lattice is embedded into a volume $\Omega$ containing $N_e$ itinerant electrons which couple to the lattice spins via on-site spin-spin interactions. The Hamiltonian for the entire system reads,

$$H = H_e + H_\mathcal{I} + J \sum_{j=1}^{N_I} \hat{\mathbf{S}}_j \cdot \hat{\mathbf{I}}_j + H_Z(\mathbf{Q}), \tag{1}$$

where $H_e = H_0 + V + U = \sum_{i=1}^{N_e} \hat{\mathbf{p}}_i^2/2m + \sum_{i<j} V_{ij} + \sum_i U(\hat{\mathbf{r}}_i)$ is the interacting electron gas Hamiltonian, $V_{ij} = V(\hat{\mathbf{r}}_i - \hat{\mathbf{r}}_j)$ the potential describing the interaction between two electrons at position $\hat{\mathbf{r}}_i$ and $\hat{\mathbf{r}}_j$, respectively, and $U(\hat{\mathbf{r}}_i)$ is an arbitrary potential for an electron at position $\hat{\mathbf{r}}_i$. Further, $J$ denotes the coupling strength of the isotropic spin interaction at lattice site $\mathbf{R}_j$, $H_J = J \sum_{j=1}^{N_I} \hat{\mathbf{S}}_j \cdot \hat{\mathbf{I}}_j$, where $\hat{\mathbf{S}}_j \equiv \hat{\mathbf{S}}(\mathbf{R}_j)$ is the electron spin density operator $\hat{\mathbf{S}}(\mathbf{r}) = \sum_{i=1}^{N_e} \hat{\mathbf{s}}_i \delta(\mathbf{r} - \hat{\mathbf{r}}_i)$ with $\hat{\mathbf{s}}_i = (\hat{s}_i^x, \hat{s}_i^y, \hat{s}_i^z)$ being the spin-1/2 of the $i^{\text{th}}$ electron. The lattice spin Hamiltonian $H_\mathcal{I} = \sum_{ij} \mathcal{I}_{ij} \hat{\mathbf{I}}_i \cdot \hat{\mathbf{I}}_j$ describes an isotropic interaction between lattice spins with coupling constants $\mathcal{I}_{ij}$ satisfying $1/N_I \sum_{ij} |\mathcal{I}_{ij}| (\mathbf{R}_i - \mathbf{R}_j)^2 < \infty$. Finally, the Zeeman term $H_Z(\mathbf{Q})$, accounts for the presence of an external (fictitious) magnetic field which breaks the symmetry of the lattice spins. To rule out (anti-)ferromagnetic order we choose $H_Z(\mathbf{Q}) = h \sum_{j=1}^{N_I} e^{-i\mathbf{Q}\cdot\mathbf{R}_j} \hat{I}_j^z + \text{h.c.}$ and to exlcude helical order we choose $\widetilde{H}_Z(\mathbf{Q}) = h \sum_{j=1}^{N_I} e^{-i\mathbf{Q}\cdot\mathbf{R}_j} \hat{I}_j^+ + \text{h.c.}$

## II. BOGOLIUBOV INEQUALITY

For the proof to rule out order we follow Ref. [1] and make use of the Bogoliubov inequality [2], which is an exact relation between two operators $A$, $C$, and a Hamiltonian $H$,

$$\frac{1}{2}\langle\{A, A^\dagger\}\rangle \langle[[C, H], C^\dagger]\rangle \geq k_B T |\langle[C, A]\rangle|^2, \tag{2}$$

where, $\langle A \rangle = Tr e^{-H/k_B T} A / Tr e^{-H/k_B T}$ denotes the expectation value in a canonical ensemble, $T$ the temperature, $k_B$ the Boltzmann constant, and $\{A, B\} = AB + BA$ the anticommutator and $[A, B] = AB - BA$ the commutator. It is assumed that all expectation values are well-defined and exist in the thermodynamic limit defined by $N_e, N_I, \Omega \to \infty$ with electron density $n_e = N_e/\Omega$ and density of lattice spins $n_I = N_I/\Omega$ finite.

If the operators $A$ and $B$ depend on the reciprocal vector $\mathbf{q}$, then the Bogoliubov inequality can be generalized to

$$\frac{1}{2}\sum_\mathbf{q} \langle\{A_\mathbf{q}, A_\mathbf{q}^\dagger\}\rangle \geq k_B T \sum_\mathbf{q} \frac{|\langle[C_\mathbf{q}, A_\mathbf{q}]\rangle|^2}{\langle[[C_\mathbf{q}, H], C_\mathbf{q}^\dagger]\rangle}, \tag{3}$$

where the sum runs over all $\mathbf{q}$'s in the first Brillouin zone of the reciprocal lattice.

## III. (ANTI-)FERROMAGNETIC ORDERING

In this case, the Zeeman term is given by $H_Z(\mathbf{Q}) = h \sum_j e^{-i\mathbf{Q}\cdot\mathbf{R}_j} \hat{I}_j^z + h.c.$ and the magnetization is defined by

$$m_I^z(\mathbf{Q}) = \frac{1}{N_I}\left\langle \sum_j \left(e^{-i\mathbf{Q}\cdot\mathbf{R}_j}\hat{I}_j^z + e^{i\mathbf{Q}\cdot\mathbf{R}_j}\hat{I}_j^z\right)\right\rangle. \tag{4}$$

We choose $C_{\mathbf{q}}$ and $A_{\mathbf{q}}$ in the Bogoliubov inequality (3) such that

$$C_{\mathbf{q}} = \underbrace{\hat{S}^-_{-\mathbf{q}} + \hat{I}^-_{-\mathbf{q}}}_{C^-_{\mathbf{q}}} + \underbrace{\hat{S}^+_{-\mathbf{q}} + \hat{I}^+_{-\mathbf{q}}}_{C^+_{\mathbf{q}}} \quad \text{and} \quad A_{\mathbf{q}} = \hat{I}^+_{\mathbf{q}+\mathbf{Q}} + I^+_{\mathbf{q}-\mathbf{Q}}. \tag{5}$$

The strategy of the proof consists in evaluating every term which enters the Bogoliubov inequality (3) to derive an upper bound for the order parameter corresponding to the phase transition we want to discuss. If this bound turns out to be in contradiction with the presence of long-range magnetic ordering, then the absence of the corresponding magnetic phase transition is rigorously demonstrated.

### A. Evaluation of $[[C_{\mathbf{q}}, H], (C_{\mathbf{q}})^\dagger]$

We decompose the double commutator $[[C_{\mathbf{q}}, H], (C_{\mathbf{q}})^\dagger]$ into the following four parts which we calculate below,

$$i)[[C^-_{\mathbf{q}} + C^+_{\mathbf{q}}, H_e], (C^-_{\mathbf{q}} + C^+_{\mathbf{q}})^\dagger], \quad ii)[[C^-_{\mathbf{q}} + C^+_{\mathbf{q}}, H_A], (C^-_{\mathbf{q}} + C^+_{\mathbf{q}})^\dagger], \quad iii)[[C^-_{\mathbf{q}} + C^+_{\mathbf{q}}, H_Z(\mathbf{Q})], (C^-_{\mathbf{q}} + C^+_{\mathbf{q}})^\dagger], \quad \text{and} \tag{6}$$

$$iv)[[C^-_{\mathbf{q}} + C^+_{\mathbf{q}}, H_\mathcal{I}]. \tag{7}$$

#### 1. Evaluation of i)

Let us first determine $[C^-_{\mathbf{q}}, H_e]$,

$$[C^-_{\mathbf{q}}, H_e] = [\hat{S}^-_{-\mathbf{q}} + \hat{I}^-_{-\mathbf{q}}, H_0 + V + U] = [\sum_i e^{i\mathbf{q}\cdot\hat{\mathbf{r}}_i} \hat{s}^-_i + \sum_j e^{i\mathbf{q}\cdot\mathbf{R}_j} \hat{I}^-_j, H_0 + V + U] = [\sum_i e^{i\mathbf{q}\cdot\hat{\mathbf{r}}_i} \hat{s}^-_i, \sum_k \hat{\mathbf{p}}^2_k/2m]$$

$$= \sum_i \frac{s^-_i}{2m}[e^{i\mathbf{q}\cdot\hat{\mathbf{r}}_i}, \hat{\mathbf{p}}^2_i] = \sum_i \frac{s^-_i}{2m}\hat{\mathbf{p}}_i \cdot [e^{i\mathbf{q}\cdot\hat{\mathbf{r}}_i}, \hat{\mathbf{p}}_i] + \sum_i \frac{s^-_i}{2m}[e^{i\mathbf{q}\cdot\hat{\mathbf{r}}_i}, \hat{\mathbf{p}}_i] \cdot \hat{\mathbf{p}}_i. \tag{8}$$

Since

$$[e^{i\mathbf{q}\cdot\hat{\mathbf{r}}_i}, \hat{\mathbf{p}}_i]\psi = (e^{i\mathbf{q}\cdot\hat{\mathbf{r}}_i}\hat{\mathbf{p}}_i - \hat{\mathbf{p}}_i e^{iq\hat{\mathbf{r}}_i})\psi = -ie^{i\mathbf{q}\cdot\hat{\mathbf{r}}_i}\psi' - \mathbf{q}e^{i\mathbf{q}\cdot\hat{\mathbf{r}}_i}\psi + ie^{i\mathbf{q}\cdot\hat{\mathbf{r}}_i}\psi' = -\mathbf{q}e^{i\mathbf{q}\cdot\hat{\mathbf{r}}_i}\psi, \tag{9}$$

we can conclude that

$$[C^-_{\mathbf{q}}, H_e] = -\mathbf{q}\sum_i \frac{s^-_i}{2m}\{\hat{\mathbf{p}}_i, e^{i\mathbf{q}\cdot\hat{\mathbf{r}}_i}\}, \tag{10}$$

and thus that

$$[[C^-_{\mathbf{q}}, H_e], (C^-_{\mathbf{q}})^\dagger] = -\frac{1}{2m}\sum_i [\hat{s}^-_i \mathbf{q} \cdot \{\hat{\mathbf{p}}_i, e^{i\mathbf{q}\cdot\hat{\mathbf{r}}_i}\}, (C^-_{\mathbf{q}})^\dagger] = -\frac{1}{2m}[\sum_i s^-_i \mathbf{q} \cdot \{\hat{\mathbf{p}}_i, e^{i\mathbf{q}\cdot\hat{\mathbf{r}}_i}\}, \sum_k \hat{s}^+_k e^{-i\mathbf{q}\cdot\hat{\mathbf{r}}_k}]$$

$$= -\frac{1}{2m}\sum_i \left( \underbrace{\hat{s}^-_i \hat{s}^+_i [\mathbf{q} \cdot \{\hat{\mathbf{p}}_i, e^{i\mathbf{q}\cdot\hat{\mathbf{r}}_i}\}, e^{-i\mathbf{q}\cdot\hat{\mathbf{r}}_i}]}_{A} + \underbrace{[\hat{s}^-_i, \hat{s}^+_i]e^{-i\mathbf{q}\cdot\hat{\mathbf{r}}_i}\mathbf{q} \cdot \{\hat{\mathbf{p}}_i, e^{i\mathbf{q}\cdot\hat{\mathbf{r}}_i}\}}_{B} \right). \tag{11}$$

We can now evaluate expressions $A$ and $B$,

$$A = (1/2 - \hat{s}^z_i)\mathbf{q} \cdot \left([\hat{\mathbf{p}}_i e^{i\mathbf{q}\cdot\hat{\mathbf{r}}_i}, e^{-i\mathbf{q}\cdot\hat{\mathbf{r}}_i}] + [e^{i\mathbf{q}\cdot\hat{\mathbf{r}}_i}\hat{\mathbf{p}}_i, e^{-i\mathbf{q}\cdot\hat{\mathbf{r}}_i}]\right) = (1/2 - \hat{s}^z_i)\mathbf{q} \cdot \left(e^{i\mathbf{q}\cdot\hat{\mathbf{r}}_i}\hat{\mathbf{p}}_i e^{-i\mathbf{q}\cdot\hat{\mathbf{r}}_i} - e^{-i\mathbf{q}\cdot\hat{\mathbf{r}}_i}\hat{\mathbf{p}}_i e^{i\mathbf{q}\cdot\hat{\mathbf{r}}_i}\right)$$

$$= (1/2 - \hat{s}^z_i)\mathbf{q} \cdot (\hat{\mathbf{p}}_i - \mathbf{q} - \hat{\mathbf{p}}_i - \mathbf{q}) = -2(1/2 - \hat{s}^z_i)q^2, \tag{12}$$

$$B = -2\hat{s}^z_i \mathbf{q} \cdot (\underbrace{e^{-i\mathbf{q}\cdot\hat{\mathbf{r}}_i}\hat{\mathbf{p}}_i e^{i\mathbf{q}\cdot\hat{\mathbf{r}}_i}}_{\hat{\mathbf{p}}_i + \mathbf{q}} + \hat{\mathbf{p}}_i) = -2\hat{s}^z_i(2\mathbf{q} \cdot \hat{\mathbf{p}}_i + q^2). \tag{13}$$

With the help of the above results, we can finally conclude that

$$[[C^-_{\mathbf{q}}, H_e], (C^-_{\mathbf{q}})^\dagger] = -\frac{1}{2m}\sum_i (1/2 - s^z_i)(-2q^2) - \frac{1}{2m}\sum_i (-2\hat{s}^z_i)(2\mathbf{q} \cdot \hat{\mathbf{p}}_i + q^2) = \frac{4\mathbf{q}}{2m}\sum_i \hat{s}^z_i\hat{\mathbf{p}}_i + \frac{1}{2m}N_e q^2. \tag{14}$$

Let us now calculate $[[C_{\mathbf{q}}^+, H_e], (C_{\mathbf{q}}^+)^\dagger]$. Since

$$[C_{\mathbf{q}}^+, H_e] = \sum_i \left( \frac{\hat{s}_i^+}{2m}\hat{\mathbf{p}}_i \cdot [e^{i\mathbf{q}\cdot\hat{\mathbf{r}}_i}, \hat{\mathbf{p}}_i] + \frac{\hat{s}_i^+}{2m}[e^{i\mathbf{q}\cdot\hat{\mathbf{r}}_i}, \hat{\mathbf{p}}_i] \cdot \hat{\mathbf{p}}_i \right) = -\mathbf{q}\cdot\sum_i \frac{\hat{s}_i^+}{2m}\{\hat{\mathbf{p}}_i, e^{i\mathbf{q}\cdot\hat{\mathbf{r}}_i}\}, \tag{15}$$

we thus have that

$$\begin{aligned}
[[C_{\mathbf{q}}^+, H_e], (C_{\mathbf{q}}^+)^\dagger] &= -\frac{1}{2m}\sum_i \hat{s}_i^+ \hat{s}_i^- [\mathbf{q}\cdot\{\hat{\mathbf{p}}_i, e^{i\mathbf{q}\cdot\hat{\mathbf{r}}_i}\}, e^{-i\mathbf{q}\cdot\hat{\mathbf{r}}_i}] - \frac{1}{2m}\sum_i [\hat{s}_i^+, \hat{s}_i^-] e^{-i\mathbf{q}\cdot\hat{\mathbf{r}}_i}\mathbf{q}\cdot\{\hat{\mathbf{p}}_i, e^{i\mathbf{q}\cdot\hat{\mathbf{r}}_i}\} \\
&= -\frac{1}{2m}\sum_i (1/2 + \hat{s}_i^z)(-2q^2) - \frac{1}{2m}\sum_i 2\hat{s}_i^z(2\mathbf{q}\cdot\hat{\mathbf{p}}_i + q^2) = \frac{1}{2m}N_e q^2 - \frac{4\mathbf{q}}{2m}\sum_i \hat{s}_i^z \hat{\mathbf{p}}_i.
\end{aligned} \tag{16}$$

Let us now calculate the mixed term $[[C_{\mathbf{q}}^+, H_e], (C_{\mathbf{q}}^-)^\dagger]$. From

$$[C_{\mathbf{q}}^+, H_e] = -\mathbf{q}\cdot\sum_i \frac{\hat{s}_i^+}{2m}\{\hat{\mathbf{p}}_i, e^{i\mathbf{q}\cdot\hat{\mathbf{r}}_i}\}, \tag{17}$$

we have that

$$\begin{aligned}
[[C_{\mathbf{q}}^+, H_e], (C_{\mathbf{q}}^-)^\dagger] &= -\frac{1}{2m}\sum_i [\hat{s}_i^+ \mathbf{q}\cdot\{\hat{\mathbf{p}}_i, e^{i\mathbf{q}\cdot\hat{\mathbf{r}}_i}\}, \hat{s}_i^+ e^{-i\mathbf{q}\cdot\hat{\mathbf{r}}_i}] \\
&= -\frac{1}{2m}\sum_i \hat{s}_i^+ \hat{s}_i^+ [\mathbf{q}\cdot\{\hat{\mathbf{p}}_i, e^{i\mathbf{q}\cdot\hat{\mathbf{r}}_i}\}, e^{-i\mathbf{q}\cdot\hat{\mathbf{r}}_i}] - \frac{1}{2m}\sum_i [\hat{s}_i^+, \hat{s}_i^+] e^{-i\mathbf{q}\cdot\hat{\mathbf{r}}_i}\mathbf{q}\cdot\{\hat{\mathbf{p}}_i, e^{i\mathbf{q}\cdot\hat{\mathbf{r}}_i}\} = 0.
\end{aligned} \tag{18}$$

An exactly analogous calculation shows that $[[C_{\mathbf{q}}^-, H_e], (C_{\mathbf{q}}^+)^\dagger] = 0$. This finally allows us to conclude that

$$[[C_{\mathbf{q}}, H], C_{\mathbf{q}}^\dagger] = \frac{1}{m}N_e q^2. \tag{19}$$

### 2. Evaluation of ii)

Let us calculate $[[C_{\mathbf{q}}^- + C_{\mathbf{q}}^+, H_J], (C_{\mathbf{q}}^- + C_{\mathbf{q}}^+)^\dagger]$. As a first step we evaluate

$$[C_{\mathbf{q}}^- + C_{\mathbf{q}}^+, H_J] = J[C_{\mathbf{q}}^- + C_{\mathbf{q}}^+, \sum_j \hat{\mathbf{S}}_j \cdot \hat{\mathbf{I}}_j]. \tag{20}$$

From

$$\begin{aligned}
[\hat{S}_{-\mathbf{q}}^-, \sum_j \hat{\mathbf{S}}_j \cdot \hat{\mathbf{I}}_j] &= [\sum_i e^{i\mathbf{q}\cdot\hat{\mathbf{r}}_i}\hat{s}_i^-, \sum_{j,k}\hat{\mathbf{s}}_k \delta(\hat{\mathbf{r}}_k - \mathbf{R}_j)\hat{\mathbf{I}}_j] = \sum_{i,j}[e^{i\mathbf{q}\hat{\mathbf{r}}_i}\hat{s}_i^-, \delta(\hat{\mathbf{r}}_i - \mathbf{R}_j)\hat{\mathbf{s}}_i \cdot \hat{\mathbf{I}}_j] \\
&= \sum_{\alpha=x,y,z}\sum_{i,j} e^{i\mathbf{q}\cdot\hat{\mathbf{r}}_i}\delta(\hat{\mathbf{r}}_i - \mathbf{R}_j)\hat{I}_j^\alpha [\hat{s}_i^-, \hat{s}_i^\alpha] = \sum_{\alpha=x,y,z}\sum_{i,j} e^{i\mathbf{q}\cdot\hat{\mathbf{r}}_i}\delta(\hat{\mathbf{r}}_i - \mathbf{R}_j)\hat{I}_j^\alpha i\hat{s}_i^\beta (\epsilon_{x\alpha\beta} - i\epsilon_{y\alpha\beta}) \\
&= \sum_{i,j} e^{i\mathbf{q}\cdot\hat{\mathbf{r}}_i}\delta(\hat{\mathbf{r}}_i - \mathbf{R}_j) i(\hat{\mathbf{I}}_j \times \hat{\mathbf{s}}_i)^-,
\end{aligned} \tag{21}$$

and

$$\begin{aligned}
[\hat{I}_{-\mathbf{q}}^-, \sum_j \hat{\mathbf{S}}_j \cdot \hat{\mathbf{I}}_j] &= [\sum_l e^{i\mathbf{q}\cdot\mathbf{R}_l}\hat{I}_l^-, \sum_{j,i}\hat{\mathbf{s}}_i \delta(\hat{\mathbf{r}}_i - \mathbf{R}_j)\hat{\mathbf{I}}_j] = \sum_{\alpha=x,y,z}\sum_{i,j} e^{i\mathbf{q}\cdot\mathbf{R}_j}\hat{s}_i^\alpha \delta(\hat{\mathbf{r}}_i - \mathbf{R}_j)[I_j^-, I_j^\alpha] \\
&= \sum_{\alpha=x,y,z}\sum_{i,j} e^{i\mathbf{q}\cdot\mathbf{R}_j}\delta(\hat{\mathbf{r}}_i - \mathbf{R}_j) i\hat{s}_i^\alpha \hat{I}_j^\beta (\epsilon_{x\alpha\beta} - i\epsilon_{y\alpha\beta}) = \sum_{i,j} e^{i\mathbf{q}\cdot\mathbf{R}_j}\delta(\hat{\mathbf{r}}_i - \mathbf{R}_j) i(\hat{\mathbf{s}}_i \times \hat{\mathbf{I}}_j)^-,
\end{aligned} \tag{22}$$

we have that $[\hat{S}_{-\mathbf{q}}^-, \sum_j \hat{\mathbf{S}}_j \cdot \hat{\mathbf{I}}_j] = -[\hat{I}_{-\mathbf{q}}^-, \sum_j \hat{\mathbf{S}}_j \cdot \hat{\mathbf{I}}_j]$ and consequently $[C_{\mathbf{q}}^-, H_J] = 0$. A similar calculation for $[C_{\mathbf{q}}^+, H_J]$ shows that

$$[C_{\mathbf{q}}^+, H_J] = J\sum_{i,j} e^{i\mathbf{q}\cdot\hat{\mathbf{r}}_i}\delta(\hat{\mathbf{r}}_i - \mathbf{R}_j) i(\hat{\mathbf{I}}_j \times \hat{\mathbf{s}}_i)^+ + J\sum_{i,j} e^{i\mathbf{q}\cdot\mathbf{R}_j}\delta(\hat{\mathbf{r}}_i - \mathbf{R}_j) i(\hat{\mathbf{s}}_i \times \hat{\mathbf{I}}_j)^+ = 0. \tag{23}$$

From this it follows that $[[C_{\mathbf{q}}^- + C_{\mathbf{q}}^+, H_J], (C_{\mathbf{q}}^- + C_{\mathbf{q}}^+)^\dagger] = 0$.

### 3. Evaluation of iii)

Let us calculate $[[C_{\mathbf{q}}, H_Z(\mathbf{Q})], C_{\mathbf{q}}^\dagger]$. From

$$\begin{aligned}
[C_{\mathbf{q}}, H_Z(\mathbf{Q})] &= [\hat{I}_{-\mathbf{q}}^- + \hat{I}_{-\mathbf{q}}^+, H_Z(\mathbf{Q})] \\
&= h\sum_j \left( e^{i\mathbf{R}_j \cdot (\mathbf{q}-\mathbf{Q})}[\hat{I}_j^-, \hat{I}_j^z] + e^{i\mathbf{R}_j \cdot (\mathbf{q}+\mathbf{Q})}[\hat{I}_j^-, \hat{I}_j^z] + e^{i\mathbf{R}_j \cdot (\mathbf{q}-\mathbf{Q})}[\hat{I}_j^+, \hat{I}_j^z] + e^{i\mathbf{R}_j \cdot (\mathbf{q}+\mathbf{Q})}[\hat{I}_j^+, \hat{I}_j^z] \right) \\
&= h\sum_j \left( e^{i\mathbf{R}_j \cdot (\mathbf{q}-\mathbf{Q})}(\hat{I}_j^- - \hat{I}_j^+) + e^{i\mathbf{R}_j \cdot (\mathbf{q}+\mathbf{Q})}(\hat{I}_j^- - \hat{I}_j^+) \right),
\end{aligned} \quad (24)$$

we obtain

$$\begin{aligned}
[[C_{\mathbf{q}}, H_Z(\mathbf{Q})], C_{\mathbf{q}}^\dagger] &= h[\sum_j \left( e^{i\mathbf{R}_j \cdot (\mathbf{q}-\mathbf{Q})}(\hat{I}_j^- - \hat{I}_j^+) + e^{i\mathbf{R}_j \cdot (\mathbf{q}+\mathbf{Q})}(\hat{I}_j^- - \hat{I}_j^+) \right), \sum_l \left( e^{-i\mathbf{q} \cdot \mathbf{R}_l}\hat{I}_l^+ + e^{-i\mathbf{q} \cdot \mathbf{R}_l}\hat{I}_l^- \right)] \\
&= \sum_j e^{-i\mathbf{Q} \cdot \mathbf{R}_j}[\hat{I}_j^-, \hat{I}_j^+] - \sum_j e^{-i\mathbf{Q} \cdot \mathbf{R}_j}[\hat{I}_j^+, \hat{I}_j^-] + \sum_j e^{i\mathbf{Q} \cdot \mathbf{R}_j}[\hat{I}_j^-, \hat{I}_j^+] - \sum_j e^{i\mathbf{Q} \cdot \mathbf{R}_j}[\hat{I}_j^+, \hat{I}_j^-] \\
&= 2h\sum_j e^{-i\mathbf{Q} \cdot \mathbf{R}_j}[\hat{I}_j^-, \hat{I}_j^+] + 2h\sum_j e^{i\mathbf{Q} \cdot \mathbf{R}_j}[\hat{I}_j^-, \hat{I}_j^+] = -4h\sum_j \left( e^{-i\mathbf{Q} \cdot \mathbf{R}_j}\hat{I}_j^z + e^{i\mathbf{Q} \cdot \mathbf{R}_j}\hat{I}_j^z \right). \quad (25)
\end{aligned}$$

### 4. Evaluation of iv)

Let us calculate $[[C_{\mathbf{q}}^- + C_{\mathbf{q}}^+, H_{\mathcal{I}}], (C_{\mathbf{q}}^- + C_{\mathbf{q}}^+)^\dagger]$. Since

$$\begin{aligned}
[C_{\mathbf{q}}^- + C_{\mathbf{q}}^+, H_{\mathcal{I}}] &= [\hat{I}_{-\mathbf{q}}^+ + \hat{I}_{-\mathbf{q}}^-, \sum_{\alpha,j,l} \mathcal{I}_{jl}\hat{I}_j^\alpha \hat{I}_l^\alpha] = [2\sum_k e^{i\mathbf{q}\cdot\mathbf{R}_k}\hat{I}_k^x, \sum_{\alpha,j,l} \mathcal{I}_{jl}\hat{I}_j^\alpha \hat{I}_l^\alpha] \\
&= 2\sum_{\alpha,k,j,l} \mathcal{I}_{jl}e^{i\mathbf{q}\cdot\mathbf{R}_k}[\hat{I}_k^x, \hat{I}_j^\alpha \hat{I}_l^\alpha] = 2\sum_{\alpha,k,j,l} \mathcal{I}_{jl}e^{i\mathbf{q}\cdot\mathbf{R}_k}\left([\hat{I}_k^x, \hat{I}_j^\alpha]\hat{I}_l^\alpha + \hat{I}_j^\alpha[\hat{I}_k^x, \hat{I}_l^\alpha]\right) \\
&= 2\sum_{\alpha,j,l} \mathcal{I}_{jl}e^{i\mathbf{q}\cdot\mathbf{R}_j}i\epsilon_{x\alpha\beta}\hat{I}_j^\beta \hat{I}_l^\alpha + 2\sum_{\alpha,j,l} \mathcal{I}_{jl}e^{i\mathbf{q}\cdot\mathbf{R}_l}\hat{I}_j^\alpha i\epsilon_{x\alpha\beta}\hat{I}_l^\beta \\
&= 2\sum_{\alpha,j,l} i\mathcal{I}_{jl}(e^{i\mathbf{q}\cdot\mathbf{R}_j} - e^{i\mathbf{q}\cdot\mathbf{R}_l})(\hat{\mathbf{I}}_l \times \hat{\mathbf{I}}_j)_x,
\end{aligned} \quad (26)$$

we obtain

$$\begin{aligned}
[[C_{\mathbf{q}}, H_{\mathcal{I}}], C_{\mathbf{q}}^\dagger] &= [2\sum_{\alpha,j,l} i\mathcal{I}_{jl}(e^{i\mathbf{q}\cdot\mathbf{R}_j} - e^{i\mathbf{q}\cdot\mathbf{R}_l})(\hat{\mathbf{I}}_l \times \hat{\mathbf{I}}_j)_x, 2\sum_k e^{-i\mathbf{q}\cdot\mathbf{R}_k}\hat{I}_k^x] \\
&= -4i\sum_{k,j,l} \mathcal{I}_{jl}e^{-i\mathbf{q}\cdot\mathbf{R}_k}(e^{i\mathbf{q}\cdot\mathbf{R}_j} - e^{i\mathbf{q}\cdot\mathbf{R}_l})[\hat{I}_k^x, (\hat{\mathbf{I}}_l \times \hat{\mathbf{I}}_j)_x] \\
&= -4i\sum_{k,j,l} \mathcal{I}_{jl}e^{-i\mathbf{q}\cdot\mathbf{R}_k}(e^{i\mathbf{q}\cdot\mathbf{R}_j} - e^{i\mathbf{q}\cdot\mathbf{R}_l})\left(\sum_\alpha \epsilon_{x\alpha\beta}\{[\hat{I}_k^x, \hat{I}_l^\alpha]\hat{I}_j^\beta + \hat{I}_l^\alpha[\hat{I}_k^x, \hat{I}_j^\beta]\}\right) \\
&= -4i\sum_{j,l} \mathcal{I}_{jl}e^{-i\mathbf{q}\cdot\mathbf{R}_l}(e^{i\mathbf{q}\cdot\mathbf{R}_j} - e^{i\mathbf{q}\cdot\mathbf{R}_l})i\sum_\alpha \epsilon_{x\alpha\beta}\epsilon_{x\alpha\gamma}\hat{I}_l^\gamma \hat{I}_j^\beta \\
&\quad -4i\sum_{j,l} \mathcal{I}_{jl}e^{-i\mathbf{q}\cdot\mathbf{R}_j}(e^{i\mathbf{q}\cdot\mathbf{R}_j} - e^{i\mathbf{q}\cdot\mathbf{R}_l})i\sum_\alpha \epsilon_{x\alpha\beta}\epsilon_{x\beta\gamma}\hat{I}_l^\alpha \hat{I}_j^\gamma \\
&= -4i\sum_{j,l} \mathcal{I}_{jl}e^{-i\mathbf{q}\cdot\mathbf{R}_l}(e^{i\mathbf{q}\cdot\mathbf{R}_j} - e^{i\mathbf{q}\cdot\mathbf{R}_l})i(\hat{I}_l^z \hat{I}_j^z + \hat{I}_l^y \hat{I}_j^y) \\
&\quad +4i\sum_{j,l} \mathcal{I}_{jl}e^{-i\mathbf{q}\cdot\mathbf{R}_j}(e^{i\mathbf{q}\cdot\mathbf{R}_j} - e^{i\mathbf{q}\cdot\mathbf{R}_l})i(\hat{I}_l^z \hat{I}_j^z + \hat{I}_l^y \hat{I}_j^y)
\end{aligned}$$

$$
\begin{aligned}
&= 4\sum_{j,l}\mathcal{I}_{jl}(e^{i\mathbf{q}\cdot(\mathbf{R}_j-\mathbf{R}_l)}-1)(\hat{I}_l^z\hat{I}_j^z+\hat{I}_l^y\hat{I}_j^y) - 4\sum_{j,l}\mathcal{I}_{jl}(1-e^{i\mathbf{q}\cdot(\mathbf{R}_l-\mathbf{R}_j)})(\hat{I}_l^z\hat{I}_j^z+\hat{I}_l^y\hat{I}_j^y) \\
&= -8\sum_{j,l}\mathcal{I}_{jl}(\hat{I}_l^z\hat{I}_j^z+\hat{I}_l^y\hat{I}_j^y) + 4\sum_{j,l}\mathcal{I}_{jl}(e^{i\mathbf{q}\cdot(\mathbf{R}_j-\mathbf{R}_l)}+e^{i\mathbf{q}\cdot(\mathbf{R}_l-\mathbf{R}_j)})(\hat{I}_l^z\hat{I}_j^z+\hat{I}_l^y\hat{I}_j^y) \\
&= -8\sum_{j,l}\mathcal{I}_{jl}[1-\cos(\mathbf{q}\cdot(\mathbf{R}_j-\mathbf{R}_l))](\hat{I}_l^z\hat{I}_j^z+\hat{I}_l^y\hat{I}_j^y).
\end{aligned}
\tag{27}
$$

The expectation value is then given by

$$
\begin{aligned}
\langle[[C_\mathbf{q},H_\mathcal{I}],C_\mathbf{q}^\dagger]\rangle &= -8\sum_{j,l}\mathcal{I}_{jl}(1-\cos(\mathbf{q}\cdot(\mathbf{R}_j-\mathbf{R}_l)))\langle\hat{I}_l^z\hat{I}_j^z+\hat{I}_l^y\hat{I}_j^y\rangle \leq 8\sum_{j,l}|\mathcal{I}_{jl}|\frac{q^2(\mathbf{R}_j-\mathbf{R}_l)^2}{2}2I^2 \\
&= 8I^2 q^2 \sum_{j,l}|\mathcal{I}_{jl}|(\mathbf{R}_j-\mathbf{R}_l)^2 = 8I^2 q^2 N_I \underbrace{\frac{1}{N_I}\sum_{j,l}|\mathcal{I}_{jl}|(\mathbf{R}_j-\mathbf{R}_l)^2}_{<\infty}.
\end{aligned}
\tag{28}
$$

### B. Evaluation of $[C_\mathbf{q}, A_\mathbf{q}]$

Let us calculate $[C_\mathbf{q}, A_\mathbf{q}]$:

$$
\begin{aligned}
[C_\mathbf{q},A_\mathbf{q}] &= [\hat{I}_{-\mathbf{q}}^- + \hat{I}_{-\mathbf{q}}^+, \hat{I}_{\mathbf{q}+\mathbf{Q}}^+ + \hat{I}_{\mathbf{q}-\mathbf{Q}}^+] = [\hat{I}_{-\mathbf{q}}^-, \hat{I}_{\mathbf{q}+\mathbf{Q}}^+ + \hat{I}_{\mathbf{q}-\mathbf{Q}}^+] = \sum_j\left(e^{-i\mathbf{Q}\cdot\mathbf{R}_j}[\hat{I}_j^-,\hat{I}_j^+]+e^{i\mathbf{Q}\cdot\mathbf{R}_j}[\hat{I}_j^-,\hat{I}_j^+]\right) \\
&= -2\sum_j\left(e^{-i\mathbf{Q}\cdot\mathbf{R}_j}\hat{I}_j^z + e^{i\mathbf{Q}\cdot\mathbf{R}_j}\hat{I}_j^z\right).
\end{aligned}
\tag{29}
$$

### C. Evaluation of $\sum_\mathbf{q}\{A_\mathbf{q}, A_\mathbf{q}^\dagger\}$

As a first step let us calculate $\{A_\mathbf{q},A_\mathbf{q}^\dagger\}$,

$$
\begin{aligned}
\{A_\mathbf{q},A_\mathbf{q}^\dagger\} &= \left\{\sum_l\left(e^{-i(\mathbf{q}+\mathbf{Q})\cdot\mathbf{R}_l}\hat{I}_l^+ + e^{-i(\mathbf{q}-\mathbf{Q})\cdot\mathbf{R}_l}\hat{I}_l^+\right), \sum_j\left(e^{i(\mathbf{q}+\mathbf{Q})\cdot\mathbf{R}_j}\hat{I}_j^- + e^{i(\mathbf{q}-\mathbf{Q})\cdot\mathbf{R}_j}\hat{I}_j^-\right)\right\} \\
&= \sum_{l,j}\left(e^{i(\mathbf{q}+\mathbf{Q})\cdot(\mathbf{R}_j-\mathbf{R}_l)}\{\hat{I}_l^+,\hat{I}_j^-\} + e^{i\mathbf{q}\cdot(\mathbf{R}_j-\mathbf{R}_l)}e^{-i\mathbf{Q}\cdot(\mathbf{R}_l+\mathbf{R}_j)}\{\hat{I}_l^+,\hat{I}_j^-\}\right) \\
&\quad + \sum_{l,j}\left(e^{i\mathbf{q}\cdot(\mathbf{R}_j-\mathbf{R}_l)}e^{i\mathbf{Q}\cdot(\mathbf{R}_l+\mathbf{R}_j)}\{\hat{I}_l^+,\hat{I}_j^-\} + e^{i(\mathbf{q}-\mathbf{Q})(\mathbf{R}_j-\mathbf{R}_l)}\{\hat{I}_l^+,\hat{I}_j^-\}\right).
\end{aligned}
\tag{30}
$$

With the use of $\sum_\mathbf{q} e^{i\mathbf{q}\cdot(\mathbf{R}_j-\mathbf{R}_l)} = N_I\delta_{\mathbf{R}_l,\mathbf{R}_j}$, we obtain

$$
\sum_\mathbf{q}\{A_\mathbf{q},A_\mathbf{q}^\dagger\} = N_I\left(2\sum_j\{\hat{I}_j^+,\hat{I}_j^-\} + \sum_j(e^{-i2\mathbf{Q}\cdot\mathbf{R}_j}+e^{i2\mathbf{Q}\cdot\mathbf{R}_j})\{\hat{I}_j^+,\hat{I}_j^-\}\right) = N_I\sum_j\{\hat{I}_j^+,\hat{I}_j^-\}(2+2\cos(\mathbf{Q}\cdot\mathbf{R}_j)),
\tag{31}
$$

and consequently

$$
\left\langle\sum_\mathbf{q}\{A_\mathbf{q},A_\mathbf{q}^\dagger\}\right\rangle \leq 4N_I\sum_j\langle\{\hat{I}_j^+,\hat{I}_j^-\}\rangle \leq 4N_I^2(2I)^2,
\tag{32}
$$

where we have used $\langle\{\hat{I}_i^+,\hat{I}_i^-\}\rangle \leq (2I)^2$. Indeed, $\{\hat{I}_i^-,\hat{I}_i^+\} = 2(\hat{I}_i^x)^2 + 2(\hat{I}_i^y)^2 \leq 4I^2$.

### D. Putting everything together

The Bogoliubov inequality (3) takes now the following form,

$$4N_I^2(2I)^2/2 \geq k_B T \sum_{\mathbf{q}} \frac{4N_I^2 m_I^z(\mathbf{Q})^2}{(\frac{N_e}{m} + 8I^2 N_I \frac{1}{N_I} \sum_{jl} |\mathcal{I}_{jl}|(\mathbf{R}_j - \mathbf{R}_l)^2)q^2 - 4hN_I m_I^z(\mathbf{Q})}. \tag{33}$$

In the thermodynamic limit we can replace the sum by an integral and obtain

$$4N_I^2(2I)^2/2 \geq \frac{k_B T N_I v}{(2\pi)^d} \int_{|\mathbf{q}| \leq |\mathbf{q}_c|} d^d q \frac{4N_I^2 m_I^z(\mathbf{Q})^2}{\frac{N_e}{m^*} q^2 - 4hN_I m_I^z(\mathbf{Q})}, \tag{34}$$

where $v = \Omega/N_I$, $d$ is the dimensionality of the system, $m^* = m/(1+8I^2 N_I/N_e m \frac{1}{N_I} \sum_{ij} |\mathcal{I}_{ij}|(\mathbf{R}_i - \mathbf{R}_j)^2)$, and $\mathbf{q}_c$ is an arbitrary cut-off vector lying in the first Brillouin zone. Since $\langle[[C_{\mathbf{q}}, H], C_{\mathbf{q}}^\dagger]\rangle \leq N_e(q^2/m^* + |2\nu h m_I^z(\mathbf{Q})|)$, inequality (34) can be simplified to

$$(2I)^2 \geq \frac{k_B T N_I v}{N_e (2\pi)^d} \int_{|\mathbf{q}| \leq |\mathbf{q}_c|} d^d q \frac{m_I^z(\mathbf{Q})^2}{\frac{q^2}{2m^*} + |\nu h m_I^z(\mathbf{Q})|}, \tag{35}$$

where $\nu = 2N_I/N_e$.

## IV. HELICAL ORDERING

In this case the symmetry-breaking Zeeman term is given by $\widetilde{H}_Z(\mathbf{Q}) = h\sqrt{2/3} \sum_j \left(e^{-i\mathbf{Q}\cdot\mathbf{R}_j} \hat{I}_j^+ + e^{i\mathbf{Q}\cdot\mathbf{R}_j} \hat{I}_j^-\right)$ and the order parameter is a helix in the $xy$-plane: $m_I^\perp(\mathbf{Q}) = \sqrt{2/3} \frac{1}{N_I} \left\langle \sum_j \left(e^{-i\mathbf{Q}\cdot\mathbf{R}_j} \hat{I}_j^+ + e^{i\mathbf{Q}\cdot\mathbf{R}_j} \hat{I}_j^-\right) \right\rangle$. The choice of operators $\widetilde{A}_{\mathbf{q}}$ and $\widetilde{C}_{\mathbf{q}}$ in the Bogoliubov inequality (3) is as follows,

$$\widetilde{C}_{\mathbf{q}} = \hat{S}_{-\mathbf{q}}^z + \hat{I}_{-\mathbf{q}}^z \quad \text{and} \quad \widetilde{A}_{\mathbf{q}} = \frac{1}{\sqrt{3}} \left(\hat{I}_{\mathbf{q}+\mathbf{Q}}^+ - \hat{I}_{-\mathbf{q}-\mathbf{Q}}^-\right). \tag{36}$$

The numerical prefactor $1/\sqrt{3}$ is chosen for later convenience.

### A. Evaluation of $[[\widetilde{C}_{\mathbf{q}}, H_e], \widetilde{C}_{\mathbf{q}}^\dagger]$

Let us first calculate

$$[\widetilde{C}_{\mathbf{q}}, H_e] = [\sum_i e^{i\mathbf{q}\cdot\hat{\mathbf{r}}_i} \hat{s}_i^z, \sum_k \hat{\mathbf{p}}_k^2/2m] = \sum_i \frac{\hat{s}_i^z}{2m} \left([e^{i\mathbf{q}\cdot\hat{\mathbf{r}}_i}, \hat{\mathbf{p}}_i] \cdot \hat{\mathbf{p}}_i + \hat{\mathbf{p}}_i \cdot [e^{i\mathbf{q}\cdot\hat{\mathbf{r}}_i}, \hat{\mathbf{p}}_i]\right). \tag{37}$$

Since $[e^{i\mathbf{q}\cdot\hat{\mathbf{r}}_i}, \hat{\mathbf{p}}_i] = -\mathbf{q} e^{i\mathbf{q}\cdot\hat{\mathbf{r}}_i}$, it follows that

$$[\widetilde{C}_{\mathbf{q}}, H_e] = -\mathbf{q} \sum_i \frac{\hat{s}_i^z}{2m} \{\hat{\mathbf{p}}_i, e^{i\mathbf{q}\cdot\hat{\mathbf{r}}_i}\}, \tag{38}$$

and consequently that

$$[[\widetilde{C}_{\mathbf{q}}, H_e], \widetilde{C}_{\mathbf{q}}^\dagger] = -\frac{1}{2m}[\sum_i \hat{s}_i^z \mathbf{q} \cdot \{\hat{\mathbf{p}}_i, e^{i\mathbf{q}\cdot\hat{\mathbf{r}}_i}\}, \sum_i e^{-i\mathbf{q}\hat{\mathbf{r}}_i} \hat{s}_i^z] = -\frac{1}{2m} \sum_i \hat{s}_i^z \hat{s}_i^z [\mathbf{q}\cdot\{\hat{\mathbf{p}}_i, e^{i\mathbf{q}\cdot\hat{\mathbf{r}}_i}\}, e^{-i\mathbf{q}\hat{\mathbf{r}}_i}]$$

$$-\frac{1}{2m} \sum_i \underbrace{[\hat{s}_i^z, \hat{s}_i^z]}_{=0} e^{-i\mathbf{q}\cdot\hat{\mathbf{r}}_i} \mathbf{q} \cdot \{\hat{\mathbf{p}}_i, e^{i\mathbf{q}\cdot\hat{\mathbf{r}}_i}\} = -\frac{1}{2m} \sum_i \frac{1}{4}(-2q^2) = \frac{1}{4m} N_e q^2. \tag{39}$$

### B. Evaluation of $[[\widetilde{C}_\mathbf{q}, \widetilde{H}_Z(\mathbf{Q})], \widetilde{C}_\mathbf{q}^\dagger]$

From

$$\begin{aligned}[][\widetilde{C}_\mathbf{q}, \widetilde{H}_Z(\mathbf{Q})] &= [\sum_j e^{i\mathbf{q}\cdot\mathbf{R}_j}\hat{I}_j^z, \sqrt{2/3}h\sum_j e^{-i\mathbf{Q}\cdot\mathbf{R}_j}\hat{I}_j^+ + \sqrt{2/3}h\sum_j e^{i\mathbf{Q}\cdot\mathbf{R}_j}\hat{I}_j^-] \\ &= \sqrt{2/3}h\sum_j e^{i(\mathbf{q}-\mathbf{Q})\cdot\mathbf{R}_j}\hat{I}_j^+ - \sqrt{2/3}h\sum_j e^{i(\mathbf{q}+\mathbf{Q})\cdot\mathbf{R}_j}\hat{I}_j^-, \end{aligned} \quad (40)$$

we obtain

$$\begin{aligned}[][[\widetilde{C}_\mathbf{q}, \widetilde{H}_z(\mathbf{Q})], \widetilde{C}_\mathbf{q}^\dagger] &= [\sqrt{2/3}h\sum_l e^{i(\mathbf{q}-\mathbf{Q})\cdot\mathbf{R}_l}\hat{I}_l^+ - \sqrt{2/3}h\sum_l e^{i(\mathbf{q}+\mathbf{Q})\cdot\mathbf{R}_l}\hat{I}_l^-, \sum_j e^{-i\mathbf{q}\cdot\mathbf{R}_j}\hat{I}_j^z] \\ &= -\sqrt{2/3}h\sum_j e^{-i\mathbf{Q}\cdot\mathbf{R}_j}\hat{I}_j^+ - \sqrt{2/3}h\sum_j e^{i\mathbf{Q}\cdot\mathbf{R}_j}\hat{I}_j^-. \end{aligned} \quad (41)$$

### C. Evaluation of $[[\widetilde{C}_\mathbf{q}, H_J], \widetilde{C}_\mathbf{q}^\dagger]$

As a first step let us calculate

$$[\widetilde{C}_\mathbf{q}, H_J] = [\hat{S}_{-\mathbf{q}}^z + \hat{I}_{-\mathbf{q}}^z, J\sum_j \hat{\mathbf{S}}_j \cdot \hat{\mathbf{I}}_j]. \quad (42)$$

Since

$$\begin{aligned}[][\hat{S}_{-\mathbf{q}}^z, \sum_j \hat{\mathbf{S}}_j \cdot \hat{\mathbf{I}}_j] &= [\sum_i e^{i\mathbf{q}\cdot\hat{\mathbf{r}}_i}\hat{s}_i^z, \sum_{j,k} \hat{\mathbf{s}}_k \delta(\hat{\mathbf{r}}_k - \hat{\mathbf{R}}_j)\hat{\mathbf{I}}_j] = \sum_{\alpha=x,y,z}\sum_{i,j} e^{i\mathbf{q}\cdot\hat{\mathbf{r}}_i}\delta(\hat{\mathbf{r}}_i - \mathbf{R}_j)\hat{I}_j^\alpha [\hat{s}_i^z, \hat{s}_i^\alpha] \\ &= \sum_{\alpha=x,y,z}\sum_{i,j} e^{i\mathbf{q}\cdot\hat{\mathbf{r}}_i}\delta(\hat{\mathbf{r}}_i - \mathbf{R}_j)i\hat{I}_j^\alpha \hat{s}_i^\beta \epsilon_{z\alpha\beta} = \sum_{i,j} e^{i\mathbf{q}\cdot\hat{\mathbf{r}}_i}\delta(\hat{\mathbf{r}}_i - \mathbf{R}_j)i(\hat{\mathbf{I}}_j \times \hat{\mathbf{s}}_i)_z, \end{aligned} \quad (43)$$

and

$$\begin{aligned}[][\hat{I}_{-\mathbf{q}}^z, \sum_j \hat{\mathbf{S}}_j \cdot \hat{\mathbf{I}}_j] &= [\sum_l e^{i\mathbf{q}\cdot\mathbf{R}_l}\hat{I}_l^z, \sum_{i,j} \hat{\mathbf{s}}_i \delta(\hat{\mathbf{r}}_i - \mathbf{R}_j)\hat{\mathbf{I}}_j] = \sum_{\alpha=x,y,z}\sum_{i,j} e^{i\mathbf{q}\cdot\mathbf{R}_i}\delta(\hat{\mathbf{r}}_i - \mathbf{R}_j)i\hat{s}_i^\alpha \hat{I}_j^\beta \epsilon_{z\alpha\beta} \\ &= \sum_{i,j} e^{i\mathbf{q}\cdot\mathbf{R}_i}\delta(\hat{\mathbf{r}}_i - \mathbf{R}_j)i(\hat{\mathbf{s}}_i \times \hat{\mathbf{I}}_j)_z, \end{aligned} \quad (44)$$

we have $[\hat{S}_{-\mathbf{q}}^z, \sum_j \hat{\mathbf{S}}_j \cdot \hat{\mathbf{I}}_j] + [\hat{I}_{-\mathbf{q}}^z, \sum_j \hat{\mathbf{S}}_j \cdot \hat{\mathbf{I}}_j] = 0$ and consequently $[\widetilde{C}_\mathbf{q}, H_J] = 0$.

### D. Evaluation of $[[\widetilde{C}_\mathbf{q}, H_\mathcal{I}], \widetilde{C}_\mathbf{q}^\dagger]$

From

$$\begin{aligned}[][\widetilde{C}_\mathbf{q}, H_\mathcal{I}] &= \sum_{\alpha,k,j,l} e^{i\mathbf{q}\cdot\mathbf{R}_k}\mathcal{I}_{jl}[\hat{I}_k^z, \hat{I}_j^\alpha \hat{I}_l^\alpha] = \sum_{\alpha,k,j,l} \mathcal{I}_{jl} e^{i\mathbf{q}\cdot\mathbf{R}_k}\left([\hat{I}_k^z, \hat{I}_j^\alpha]\hat{I}_l^\alpha + \hat{I}_j^\alpha[\hat{I}_k^z, \hat{I}_l^\alpha]\right) \\ &= \sum_{\alpha,j,l} \mathcal{I}_{jl} e^{i\mathbf{q}\cdot\mathbf{R}_j} i\epsilon_{z\alpha\beta}\hat{I}_j^\beta \hat{I}_l^\alpha + \sum_{\alpha,j,l} \mathcal{I}_{jl} e^{i\mathbf{q}\cdot\mathbf{R}_l} \hat{I}_j^\alpha i\epsilon_{z\alpha\beta}\hat{I}_l^\beta = i\sum_{j,l} \mathcal{I}_{jl}(e^{i\mathbf{q}\cdot\mathbf{R}_j} - e^{i\mathbf{q}\cdot\mathbf{R}_l})(\hat{\mathbf{I}}_l \times \hat{\mathbf{I}}_j)_z, \end{aligned} \quad (45)$$

we obtain

$$\begin{aligned}[][[\widetilde{C}_\mathbf{q}, H_\mathcal{I}], \widetilde{C}_\mathbf{q}^\dagger] &= -i\sum_{k,j,l} \mathcal{I}_{jl} e^{-i\mathbf{q}\cdot\mathbf{R}_k}(e^{i\mathbf{q}\cdot\mathbf{R}_j} - e^{i\mathbf{q}\cdot\mathbf{R}_l})[\hat{I}_k^z, (\hat{\mathbf{I}}_l \times \hat{\mathbf{I}}_j)_z] \\ &= -i\sum_{jl} \mathcal{I}_{jl} e^{-i\mathbf{q}\cdot\mathbf{R}_l}(e^{i\mathbf{q}\cdot\mathbf{R}_j} - e^{i\mathbf{q}\cdot\mathbf{R}_l})\sum_\alpha i\epsilon_{z\alpha\beta}\epsilon_{z\alpha\gamma}\hat{I}_l^\gamma \hat{I}_j^\beta \\ &\quad -i\sum_{jl} \mathcal{I}_{jl} e^{-i\mathbf{q}\cdot\mathbf{R}_j}(e^{i\mathbf{q}\cdot\mathbf{R}_j} - e^{i\mathbf{q}\cdot\mathbf{R}_l})\sum_\alpha i\epsilon_{z\alpha\beta}\epsilon_{z\beta\gamma}\hat{I}_l^\alpha \hat{I}_j^\gamma \\ &= -2\sum_{jl} \mathcal{I}_{jl}(1 - \cos(\mathbf{q}\cdot(\mathbf{R}_j - \mathbf{R}_l)))(\hat{I}_j^x \hat{I}_l^x + \hat{I}_j^y \hat{I}_l^y). \end{aligned} \quad (46)$$

The expectation value is then given by

$$\langle[[\widetilde{C}_{\mathbf{q}}, H_{\mathcal{I}}], \widetilde{C}_{\mathbf{q}}^{\dagger}]\rangle = -2\sum_{jl}\mathcal{I}_{jl}(1-\cos(\mathbf{q}\cdot(\mathbf{R}_j-\mathbf{R}_l)))\left\langle(\hat{I}_j^x\hat{I}_l^x+\hat{I}_j^y\hat{I}_l^y)\right\rangle \leq 2\sum_{jl}|\mathcal{I}_{jl}|\frac{q^2(\mathbf{R}_j-\mathbf{R}_l)^2}{2}2I^2$$

$$= 2q^2I^2\sum_{jl}|\mathcal{I}_{jl}|(\mathbf{R}_j-\mathbf{R}_l)^2 = 2q^2N_II^2\frac{1}{N_I}\sum_{j,l}|\mathcal{I}_{jl}|(\mathbf{R}_j-\mathbf{R}_l)^2. \quad (47)$$

### E. Evaluation of $[\widetilde{C}_{\mathbf{q}}, \widetilde{A}_{\mathbf{q}}]$

Let us now calculate $[\widetilde{C}_{\mathbf{q}}, \widetilde{A}_{\mathbf{q}}]$,

$$[\widetilde{C}_{\mathbf{q}}, \widetilde{A}_{\mathbf{q}}] = [\sum_l e^{i\mathbf{q}\cdot\mathbf{R}_l}\hat{I}_l^z, \frac{1}{\sqrt{3}}\sum_j e^{-i(\mathbf{q}+\mathbf{Q})\cdot\mathbf{R}_j}\hat{I}_j^+ - \frac{1}{\sqrt{3}}\sum_j e^{-i(\mathbf{q}-\mathbf{Q})\cdot\mathbf{R}_j}\hat{I}_j^-] = \frac{1}{\sqrt{3}}\sum_j e^{-i\mathbf{Q}\cdot\mathbf{R}_j}\hat{I}_j^+ + \frac{1}{\sqrt{3}}\sum_j e^{i\mathbf{Q}\cdot\mathbf{R}_j}\hat{I}_j^-. \quad (48)$$

### F. Evaluation of $\sum_{\mathbf{q}}\{\widetilde{A}_{\mathbf{q}}, \widetilde{A}_{\mathbf{q}}^{\dagger}\}$

Let us first calculate $\{\widetilde{A}_{\mathbf{q}}, \widetilde{A}_{\mathbf{q}}^{\dagger}\}$,

$$\{\widetilde{A}_{\mathbf{q}}, \widetilde{A}_{\mathbf{q}}^{\dagger}\} = \left\{\frac{1}{\sqrt{3}}\left(\hat{I}_{\mathbf{q}+\mathbf{Q}}^+ - \hat{I}_{\mathbf{q}-\mathbf{Q}}^-\right), \frac{1}{\sqrt{3}}\left(\hat{I}_{-\mathbf{q}-\mathbf{Q}}^- - \hat{I}_{-\mathbf{q}+\mathbf{Q}}^+\right)\right\}$$

$$= \left\{\frac{1}{\sqrt{3}}\left(\sum_l e^{-i(\mathbf{q}+\mathbf{Q})\cdot\mathbf{R}_l}\hat{I}_l^+ - \sum_l e^{-i(\mathbf{q}-\mathbf{Q})\cdot\mathbf{R}_l}\hat{I}_l^-\right), \frac{1}{\sqrt{3}}\left(\sum_j e^{i(\mathbf{q}+\mathbf{Q})\cdot\mathbf{R}_j}\hat{I}_j^- - \sum_j e^{i(\mathbf{q}-\mathbf{Q})\cdot\mathbf{R}_j}\hat{I}_j^+\right)\right\}$$

$$= \frac{1}{3}\sum_{l,j}e^{i(\mathbf{q}+\mathbf{Q})\cdot(\mathbf{R}_j-\mathbf{R}_l)}\{\hat{I}_l^+,\hat{I}_j^-\} - \frac{1}{3}\sum_{l,j}e^{i\mathbf{q}(\mathbf{R}_j-\mathbf{R}_l)}e^{-i\mathbf{Q}\cdot(\mathbf{R}_j+\mathbf{R}_l)}\{\hat{I}_l^+,\hat{I}_j^+\}$$

$$-\frac{1}{3}\sum_{l,j}e^{i\mathbf{q}\cdot(\mathbf{R}_j-\mathbf{R}_l)}e^{i\mathbf{Q}\cdot(\mathbf{R}_l+\mathbf{R}_j)}\{\hat{I}_l^-,\hat{I}_j^-\} + \frac{1}{3}\sum_{l,j}e^{i(\mathbf{q}-\mathbf{Q})\cdot(\mathbf{R}_j-\mathbf{R}_l)}\{\hat{I}_l^-,\hat{I}_j^+\}. \quad (49)$$

By using $\sum_{\mathbf{q}}e^{i\mathbf{q}\cdot(\mathbf{R}_j-\mathbf{R}_l)} = N_I\delta_{\mathbf{R}_l,\mathbf{R}_j}$, we obtain

$$\sum_{\mathbf{q}}\{\widetilde{A}_{\mathbf{q}}, \widetilde{A}_{\mathbf{q}}^{\dagger}\} = \frac{1}{3}N_I\left(\sum_j\{\hat{I}_j^+,\hat{I}_j^-\} - \sum_j e^{-i2\mathbf{Q}\cdot\mathbf{R}_j}\{\hat{I}_j^+,\hat{I}_j^+\} - \sum_j e^{i2\mathbf{Q}\cdot\mathbf{R}_j}\{\hat{I}_j^-,\hat{I}_j^-\} + \sum_j\{\hat{I}_j^-,\hat{I}_j^+\}\right)$$

$$= \frac{1}{3}N_I\left(2\sum_j\{\hat{I}_j^+,\hat{I}_j^-\} - \sum_j e^{-i2\mathbf{Q}\cdot\mathbf{R}_j}\{\hat{I}_j^+,\hat{I}_j^+\} - \sum_j e^{i2\mathbf{Q}\cdot\mathbf{R}_j}\{\hat{I}_j^-,\hat{I}_j^-\}\right)$$

$$= \frac{1}{3}N_I\left(2\sum_j\{\hat{I}_j^+,\hat{I}_j^-\} + 2\sum_j((\hat{I}_j^y)^2-(\hat{I}_j^x)^2)2\cos(2\mathbf{Q}\cdot\mathbf{R}_j) - 2\sum_j\{\hat{I}_j^x,\hat{I}_j^y\}2\sin(2\mathbf{Q}\cdot\mathbf{R}_j)\right)$$

$$\leq \frac{1}{3}N_I\left(2\sum_j\{\hat{I}_j^+,\hat{I}_j^-\} + 4\sum_j((\hat{I}_j^y)^2-(\hat{I}_j^x)^2) + 4\sum_j\{\hat{I}_j^x,\hat{I}_j^y\}\right) \leq 2N_I^2(2I)^2, \quad (50)$$

where we have used that

$$\{I_j^{\pm}, I_j^{\pm}\} = 2((I_j^x)^2-(I_j^y)^2 \pm i\{I_j^x,I_j^y\}), \quad (51)$$

$$\langle(I_j^y)^2-(I_j^x)^2\rangle \leq 2I^2, \quad (52)$$

$$\langle\{I_j^x,I_j^y\}\rangle \leq 2I^2, \quad (53)$$

$$\langle\{I_j^+,I_j^-\}\rangle \leq (2I)^2. \quad (54)$$

### G. Putting everything together

With the use of the above results the Bogoliubov inequality (3) reads as follows,

$$2N_I^2(2I)^2/2 \geq k_B T \sum_\mathbf{q} \frac{N_I^2 m_I^\perp(\mathbf{Q})^2/2}{\frac{N_e}{4m^*}q^2 - hN_I m_I^\perp(\mathbf{Q})}, \tag{55}$$

where $m^*$ is defined as above. In the thermodynamic limit we thus obtain

$$(2I)^2 \geq \frac{k_B T N_I v}{N_e (2\pi)^d} \int_{|\mathbf{q}| \leq |\mathbf{q}_c|} d^d q \frac{m_I^\perp(\mathbf{Q})^2}{\frac{q^2}{2m^*} + |\nu h m_I^\perp(\mathbf{Q})|}, \tag{56}$$

where $v$ and $\nu$ are defined as above.

## V. PRESENCE OF SPIN-ORBIT INTERACTION

In this section we want to investigate the effect of the presence of Rashba and Dresselhaus spin-orbit terms which break the rotational spin symmetry of Hamiltonian (1). The spin-orbit Hamiltonian under consideration is thus given by

$$H_{\text{SO}} = H_R + H_D = \alpha \sum_i (\hat{p}_i^y \hat{s}_i^x - \hat{p}_i^x \hat{s}_i^y) + \beta \sum_i (\hat{p}_i^x \hat{s}_i^x - \hat{p}_i^y \hat{s}_i^y), \tag{57}$$

where $H_R$ and $H_D$ are the Rashba and the Dresselhaus spin-orbit Hamiltonian, respectively [3, 4].

### A. (Anti-)ferromagnetic ordering

In this case, the operator $C_\mathbf{q}$ entering the Bogoliubov inequality (3) is given by $C_\mathbf{q} = \hat{S}_{-\mathbf{q}}^- + \hat{I}_{-\mathbf{q}}^- + \hat{S}_{-\mathbf{q}}^+ + \hat{I}_{-\mathbf{q}}^+$.
Let us first calculate $[[C_\mathbf{q}, H_R], (C_\mathbf{q})^\dagger]$. From

$$\begin{aligned} [C_\mathbf{q}, H_R] &= [\hat{S}_{-\mathbf{q}}^- + \hat{S}_{-\mathbf{q}}^+, H_R] = [\sum_i e^{i\mathbf{q}\cdot\hat{\mathbf{r}}_i}(\hat{s}_i^+ + \hat{s}_i^-), \alpha \sum_i (\hat{p}_i^y \hat{s}_i^x - \hat{p}_i^x \hat{s}_i^y)] \\ &= \alpha \sum_i \left( (\hat{s}_i^+ + \hat{s}_i^-)\hat{s}_i^x [e^{i\mathbf{q}\cdot\hat{\mathbf{r}}_i}, \hat{p}_i^y] + \underbrace{[\hat{s}_i^+ + \hat{s}_i^-, \hat{s}_i^x]}_{=0} \hat{p}_i^y e^{i\mathbf{q}\hat{\mathbf{r}}_i} - (\hat{s}_i^+ + \hat{s}_i^-)\hat{s}_i^y [e^{i\mathbf{q}\cdot\hat{\mathbf{r}}_i}, \hat{p}_i^x] - [\hat{s}_i^+ + \hat{s}_i^-, \hat{s}_i^y]\hat{p}_i^x e^{i\mathbf{q}\cdot\hat{\mathbf{r}}_i} \right), \end{aligned} \tag{58}$$

and with the aid of

$$\begin{aligned} (\hat{s}_i^+ + \hat{s}_i^-)\hat{s}_i^x &= 2\hat{s}_i^x \hat{s}_i^x = \frac{1}{2}, \tag{59} \\ (\hat{s}_i^+ + \hat{s}_i^-)\hat{s}_i^y &= 2\hat{s}_i^x \hat{s}_i^y = i\hat{s}_i^z, \tag{60} \\ [\hat{s}_i^+ + \hat{s}_i^-, \hat{s}_i^y] &= [2\hat{s}_i^x, \hat{s}_i^y] = 2i\hat{s}_i^z, \tag{61} \end{aligned}$$

we have that

$$[C_\mathbf{q}, H_R] = \alpha \sum_i \left( \frac{1}{2}[e^{i\mathbf{q}\hat{\mathbf{r}}_i}, \hat{p}_i^y] - i\hat{s}_i^z [e^{i\mathbf{q}\cdot\hat{\mathbf{r}}_i}, \hat{p}_i^x] - 2i\hat{s}_i^z \hat{p}_i^x e^{i\mathbf{q}\cdot\hat{\mathbf{r}}_i} \right) = \alpha \sum_i \left( \frac{1}{2}[e^{i\mathbf{q}\cdot\hat{\mathbf{r}}_i}, \hat{p}_i^y] - i\hat{s}_i^z \{\hat{p}_i^x, e^{i\mathbf{q}\cdot\hat{\mathbf{r}}_i}\} \right). \tag{62}$$

Therefore,

$$\begin{aligned}[][[C_\mathbf{q}, H_R], C_\mathbf{q}^\dagger] &= [\alpha \sum_i \left(\frac{1}{2}[e^{i\mathbf{q}\cdot\hat{\mathbf{r}}_i}, \hat{p}_i^y] - i\hat{s}_i^z\{\hat{p}_i^x, e^{i\mathbf{q}\cdot\hat{\mathbf{r}}_i}\}\right), \sum_i e^{-i\mathbf{q}\cdot\hat{\mathbf{r}}_i}(\hat{s}_i^+ + \hat{s}_i^-)] \\
&= \alpha \sum_i (\hat{s}_i^+ + \hat{s}_i^-) \underbrace{[\frac{1}{2}[e^{i\mathbf{q}\cdot\hat{\mathbf{r}}_i}, \hat{p}_i^y], e^{-i\mathbf{q}\cdot\hat{\mathbf{r}}_i}]}_{=0} - \alpha \sum_i \underbrace{i\hat{s}_i^z(\hat{s}_i^+ + \hat{s}_i^-)}_{-\hat{s}_i^y} \underbrace{[\{\hat{p}_i^x, e^{i\mathbf{q}\cdot\hat{\mathbf{r}}_i}\}, e^{-i\mathbf{q}\cdot\hat{\mathbf{r}}_i}]}_{-2q_x} \\
&\quad - \alpha \sum_i \underbrace{i[\hat{s}_i^z, \hat{s}_i^+ + \hat{s}_i^-]}_{-2\hat{s}_i^y} e^{-i\mathbf{q}\cdot\hat{\mathbf{r}}_i} \underbrace{\{\hat{p}_i^x, e^{i\mathbf{q}\cdot\hat{\mathbf{r}}_i}\}}_{2\hat{p}_i^x + q_x} \\
&= \alpha \sum_i 4\hat{p}_i^x \hat{s}_i^y.
\end{aligned} \tag{63}$$

Let us now calculate $[[C_\mathbf{q}, H_D], (C_\mathbf{q})^\dagger]$. Analogously to the above case we have that

$$\begin{aligned}[][C_\mathbf{q}, H_D] &= [\hat{S}_{-\mathbf{q}}^- + \hat{S}_{-\mathbf{q}}^+, H_D] = [\sum_i e^{i\mathbf{q}\cdot\hat{\mathbf{r}}_i}(\hat{s}_i^+ + \hat{s}_i^-), \beta \sum_i (\hat{p}_i^x \hat{s}_i^x - \hat{p}_i^y \hat{s}_i^y)] \\
&= \beta \sum_i \left((\hat{s}_i^+ + \hat{s}_i^-)\hat{s}_i^x [e^{i\mathbf{q}\cdot\hat{\mathbf{r}}_i}, \hat{p}_i^x] + \underbrace{[\hat{s}_i^+ + \hat{s}_i^-, \hat{s}_i^x]}_{=0} \hat{p}_i^x e^{i\mathbf{q}\cdot\hat{\mathbf{r}}_i} - (\hat{s}_i^+ + \hat{s}_i^-)\hat{s}_i^y [e^{i\mathbf{q}\cdot\hat{\mathbf{r}}_i}, \hat{p}_i^y] - [\hat{s}_i^+ + \hat{s}_i^-, \hat{s}_i^y]\hat{p}_i^y e^{i\mathbf{q}\hat{\mathbf{r}}_i}\right) \\
&= \beta \sum_i \left(\frac{1}{2}[e^{i\mathbf{q}\cdot\hat{\mathbf{r}}_i}, \hat{p}_i^x] - i\hat{s}_i^z[e^{i\mathbf{q}\cdot\hat{\mathbf{r}}_i}, \hat{p}_i^y] - 2i\hat{s}_i^z \hat{p}_i^y e^{i\mathbf{q}\cdot\hat{\mathbf{r}}_i}\right) = \beta \sum_i \left(\frac{1}{2}[e^{i\mathbf{q}\cdot\hat{\mathbf{r}}_i}, \hat{p}_i^x] - i\hat{s}_i^z \{\hat{p}_i^y, e^{i\mathbf{q}\cdot\hat{\mathbf{r}}_i}\}\right),
\end{aligned} \tag{64}$$

and consequently we obtain

$$\begin{aligned}[][[C_\mathbf{q}, H_D], C_\mathbf{q}^\dagger] &= [\beta \sum_i \left(\frac{1}{2}[e^{i\mathbf{q}\hat{\mathbf{r}}_i}, \hat{p}_i^x] - i\hat{s}_i^z \{\hat{p}_i^y, e^{i\mathbf{q}\cdot\hat{\mathbf{r}}_i}\}\right), \sum_i e^{-i\mathbf{q}\cdot\hat{\mathbf{r}}_i}(\hat{s}_i^+ + \hat{s}_i^-)] \\
&= \beta \sum_i (\hat{s}_i^+ + \hat{s}_i^-) \underbrace{[\frac{1}{2}[e^{i\mathbf{q}\cdot\hat{\mathbf{r}}_i}, \hat{p}_i^x], e^{-i\mathbf{q}\hat{\mathbf{r}}_i}]}_{=0} - \beta \sum_i \underbrace{i\hat{s}_i^z(\hat{s}_i^+ + \hat{s}_i^-)}_{-\hat{s}_i^y} \underbrace{[\{\hat{p}_i^y, e^{i\mathbf{q}\cdot\hat{\mathbf{r}}_i}\}, e^{-i\mathbf{q}\cdot\hat{\mathbf{r}}_i}]}_{-2q_y} \\
&\quad - \beta \sum_i \underbrace{i[\hat{s}_i^z, \hat{s}_i^+ + \hat{s}_i^-]}_{-2\hat{s}_i^y} e^{-i\mathbf{q}\cdot\hat{\mathbf{r}}_i} \underbrace{\{\hat{p}_i^y, e^{i\mathbf{q}\cdot\hat{\mathbf{r}}_i}\}}_{2\hat{p}_i^y + q_y} \\
&= \beta \sum_i 4\hat{p}_i^y \hat{s}_i^y.
\end{aligned} \tag{65}$$

We thus finally conclude that

$$[[C_\mathbf{q}, H_{\text{SO}}], C_\mathbf{q}^\dagger] = 4 \sum_i (\alpha \hat{p}_i^x + \beta \hat{p}_i^y) \hat{s}_i^y. \tag{66}$$

### B. Helical ordering

In this case, $\widetilde{C}_\mathbf{q}$ is chosen to be $\widetilde{C}_\mathbf{q} = \hat{S}_{-\mathbf{q}}^z + \hat{I}_{-\mathbf{q}}^z$. From

$$\begin{aligned}[][\widetilde{C}_\mathbf{q}, H_R] &= [\sum_i e^{i\mathbf{q}\cdot\hat{\mathbf{r}}_i}\hat{s}_i^z, \alpha \sum_i (\hat{p}_i^y \hat{s}_i^x - \hat{p}_i^x \hat{s}_i^y)] = \alpha \sum_i \left([e^{i\mathbf{q}\cdot\hat{\mathbf{r}}_i}\hat{s}_i^z, \hat{p}_i^y \hat{s}_i^x] - [e^{i\mathbf{q}\cdot\hat{\mathbf{r}}_i}\hat{s}_i^z, \hat{p}_i^x \hat{s}_i^y]\right) \\
&= \alpha \sum_i \left(\underbrace{\hat{s}_i^z \hat{s}_i^x}_{\frac{i}{2}\hat{s}_i^y}[e^{i\mathbf{q}\cdot\hat{\mathbf{r}}_i}, \hat{p}_i^y] + \underbrace{[\hat{s}_i^z, \hat{s}_i^x]}_{i\hat{s}_i^y}\hat{p}_i^y e^{i\mathbf{q}\cdot\hat{\mathbf{r}}_i} - \underbrace{\hat{s}_i^z \hat{s}_i^y}_{-\frac{i}{2}\hat{s}_i^x}[e^{i\mathbf{q}\cdot\hat{\mathbf{r}}_i}, \hat{p}_i^x] - \underbrace{[\hat{s}_i^z, \hat{s}_i^y]}_{-i\hat{s}_i^x}\hat{p}_i^x e^{i\mathbf{q}\hat{\mathbf{r}}_i}\right) \\
&= \alpha \sum_i \left(\frac{1}{2}i\hat{s}_i^y \{\hat{p}_i^y, e^{i\mathbf{q}\hat{\mathbf{r}}_i}\} + \frac{1}{2}i\hat{s}_i^x \{\hat{p}_i^x, e^{i\mathbf{q}\cdot\hat{\mathbf{r}}_i}\}\right),
\end{aligned} \tag{67}$$

we obtain

$$
\begin{aligned}
{[[\widetilde{C}_{\mathbf{q}}, H_R], \widetilde{C}_{\mathbf{q}}^\dagger]} &= [\alpha \sum_i \left(\frac{1}{2} i \hat{s}_i^y \{\hat{p}_i^y, e^{i\mathbf{q}\cdot\hat{\mathbf{r}}_i}\} + \frac{1}{2} i \hat{s}_i^x \{\hat{p}_i^x, e^{i\mathbf{q}\cdot\hat{\mathbf{r}}_i}\}\right), \sum_i e^{-i\mathbf{q}\cdot\hat{\mathbf{r}}_i} \hat{s}_i^z] \\
&= \alpha \sum_i \left(\frac{1}{2} i [\hat{s}_i^y \{\hat{p}_i^y, e^{i\mathbf{q}\cdot\hat{\mathbf{r}}_i}\}, e^{-i\mathbf{q}\cdot\hat{\mathbf{r}}_i} \hat{s}_i^z] + \frac{1}{2} i [\hat{s}_i^x \{\hat{p}_i^x, e^{i\mathbf{q}\cdot\hat{\mathbf{r}}_i}\}, e^{-i\mathbf{q}\cdot\hat{\mathbf{r}}_i} \hat{s}_i^z]\right) \\
&= \alpha \sum_i \frac{1}{2} i \left( \underbrace{\hat{s}_i^y \hat{s}_i^z}_{\frac{i}{2}\hat{s}_i^x} \underbrace{[\{\hat{p}_i^y, e^{i\mathbf{q}\cdot\hat{\mathbf{r}}_i}\}, e^{-i\mathbf{q}\cdot\hat{\mathbf{r}}_i}]}_{-2q_y} + \underbrace{[\hat{s}_i^y, \hat{s}_i^z]}_{i\hat{s}_i^x} \underbrace{e^{-i\mathbf{q}\cdot\hat{\mathbf{r}}_i}\{\hat{p}_i^y, e^{i\mathbf{q}\cdot\hat{\mathbf{r}}_i}\}}_{2\hat{p}_i^y + q_y} \right) \\
&\quad + \alpha \sum_i \frac{1}{2} i \left( \underbrace{\hat{s}_i^x \hat{s}_i^z}_{-\frac{i}{2}\hat{s}_i^y} \underbrace{[\{\hat{p}_i^x, e^{i\mathbf{q}\cdot\hat{\mathbf{r}}_i}\}, e^{-i\mathbf{q}\cdot\hat{\mathbf{r}}_i}]}_{-2q_x} + \underbrace{[\hat{s}_i^x, \hat{s}_i^z]}_{-i\hat{s}_i^y} \underbrace{e^{-i\mathbf{q}\cdot\hat{\mathbf{r}}_i}\{\hat{p}_i^x, e^{i\mathbf{q}\cdot\hat{\mathbf{r}}_i}\}}_{2\hat{p}_i^x + q_x} \right) \\
&= \alpha \sum_i (\hat{p}_i^x \hat{s}_i^y - \hat{p}_i^y \hat{s}_i^x) = -H_R.
\end{aligned}
\tag{68}
$$

Similarly, from

$$
\begin{aligned}
{[\widetilde{C}_{\mathbf{q}}, H_D]} &= [\sum_i e^{i\mathbf{q}\hat{\mathbf{r}}_i} \hat{s}_i^z, \beta \sum_i (\hat{p}_i^x \hat{s}_i^x - \hat{p}_i^y \hat{s}_i^y)] = \beta \sum_i \left([e^{i\mathbf{q}\hat{\mathbf{r}}_i} \hat{s}_i^z, \hat{p}_i^x \hat{s}_i^x] - [e^{i\mathbf{q}\hat{\mathbf{r}}_i} \hat{s}_i^z, \hat{p}_i^y \hat{s}_i^y]\right) \\
&= \beta \sum_i \left( \underbrace{\hat{s}_i^z \hat{s}_i^x}_{\frac{i}{2}\hat{s}_i^y} [e^{i\mathbf{q}\cdot\hat{\mathbf{r}}_i}, \hat{p}_i^x] + \underbrace{[\hat{s}_i^z, \hat{s}_i^x]}_{i\hat{s}_i^y} \hat{p}_i^x e^{i\mathbf{q}\cdot\hat{\mathbf{r}}_i} - \underbrace{\hat{s}_i^z \hat{s}_i^y}_{-\frac{i}{2}\hat{s}_i^x} [e^{i\mathbf{q}\cdot\hat{\mathbf{r}}_i}, \hat{p}_i^y] - \underbrace{[\hat{s}_i^z, \hat{s}_i^y]}_{-i\hat{s}_i^x} \hat{p}_i^y e^{i\mathbf{q}\cdot\hat{\mathbf{r}}_i} \right) \\
&= \beta \sum_i \left(\frac{i}{2} \hat{s}_i^y \{\hat{p}_i^x, e^{i\mathbf{q}\cdot\hat{\mathbf{r}}_i}\} + \frac{i}{2} \hat{s}_i^x \{\hat{p}_i^y, e^{i\mathbf{q}\cdot\hat{\mathbf{r}}_i}\}\right),
\end{aligned}
\tag{69}
$$

we obtain

$$
\begin{aligned}
{[[\widetilde{C}_{\mathbf{q}}, H_D], \widetilde{C}_{\mathbf{q}}^\dagger]} &= [\beta \sum_i \left(\frac{i}{2} \hat{s}_i^y \{\hat{p}_i^x, e^{i\mathbf{q}\cdot\hat{\mathbf{r}}_i}\} + \frac{i}{2} \hat{s}_i^x \{\hat{p}_i^y, e^{i\mathbf{q}\cdot\hat{\mathbf{r}}_i}\}\right), \sum_i e^{-i\mathbf{q}\cdot\hat{\mathbf{r}}_i} \hat{s}_i^z] \\
&= \beta \sum_i \left(\frac{i}{2} [\hat{s}_i^y \{\hat{p}_i^x, e^{i\mathbf{q}\cdot\hat{\mathbf{r}}_i}\}, e^{-i\mathbf{q}\cdot\hat{\mathbf{r}}_i} \hat{s}_i^z] + \frac{i}{2} [\hat{s}_i^x \{\hat{p}_i^y, e^{i\mathbf{q}\cdot\hat{\mathbf{r}}_i}\}, e^{-i\mathbf{q}\cdot\hat{\mathbf{r}}_i} \hat{s}_i^z]\right) \\
&= \beta \sum_i \left(\frac{i}{2} \underbrace{\hat{s}_i^y \hat{s}_i^z}_{\frac{i}{2}\hat{s}_i^x} \underbrace{[\{\hat{p}_i^x, e^{i\mathbf{q}\cdot\hat{\mathbf{r}}_i}\}, e^{-i\mathbf{q}\cdot\hat{\mathbf{r}}_i}]}_{-2q_x} + \frac{i}{2} \underbrace{[\hat{s}_i^y, \hat{s}_i^z]}_{i\hat{s}_i^x} \underbrace{e^{-i\mathbf{q}\cdot\hat{\mathbf{r}}_i}\{\hat{p}_i^x, e^{i\mathbf{q}\cdot\hat{\mathbf{r}}_i}\}}_{2\hat{p}_i^x + q_x} \right) \\
&\quad + \beta \sum_i \left(\frac{i}{2} \underbrace{\hat{s}_i^x \hat{s}_i^z}_{-\frac{i}{2}\hat{s}_i^y} \underbrace{[\{\hat{p}_i^y, e^{i\mathbf{q}\cdot\hat{\mathbf{r}}_i}\}, e^{-i\mathbf{q}\cdot\hat{\mathbf{r}}_i}]}_{-2q_y} + \frac{i}{2} \underbrace{[\hat{s}_i^x, \hat{s}_i^z]}_{-i\hat{s}_i^y} \underbrace{e^{-i\mathbf{q}\cdot\hat{\mathbf{r}}_i}\{\hat{p}_i^y, e^{i\mathbf{q}\cdot\hat{\mathbf{r}}_i}\}}_{2\hat{p}_i^y + q_y} \right) \\
&= \beta \sum_i (\hat{p}_i^y \hat{s}_i^y - \hat{p}_i^x \hat{s}_i^x) = -H_D.
\end{aligned}
\tag{70}
$$

We can thus finally conclude that

$$
[[\widetilde{C}_{\mathbf{q}}, H_{\text{SO}}], \widetilde{C}_{\mathbf{q}}^\dagger] = -H_{\text{SO}}.
\tag{71}
$$

## VI. CONTINUITY EQUATION FOR SPIN-CURRENTS

Similar to Refs. [6, 7], we can derive a continuity equation for the total spin density operator $\hat{\boldsymbol{\Sigma}}(\mathbf{r}) = \hat{\mathbf{S}}(\mathbf{r}) + \hat{\mathbf{I}}(\mathbf{r})$, where the lattice spin density is defined as $\hat{\mathbf{I}}(\mathbf{r}) = \sum_{j=1}^{N_I} \hat{\mathbf{I}}_j \delta(\mathbf{r} - \mathbf{R}_j)$. Let us first recall that $\hat{S}^\alpha(\mathbf{r})$ satisfies the

Born-von Karman periodic boundary conditions and can thus be expanded as a Fourier series,

$$\hat{S}^\alpha(\mathbf{r}) = \frac{1}{\Omega} \sum_{\mathbf{q}} e^{i\mathbf{q}\cdot\mathbf{r}} \hat{S}^\alpha_{\mathbf{q}}, \tag{72}$$

where $\hat{S}^\alpha_{\mathbf{q}} = \sum_i e^{-i\mathbf{q}\cdot\hat{\mathbf{r}}_i} \hat{s}^\alpha_i$.

The Heisenberg equation of motion for the component $\alpha = x, y, z$ of the electron spin density is given by

$$\dot{\hat{S}}^\alpha(\mathbf{r}) = i[H, \hat{S}^\alpha(\mathbf{r})]. \tag{73}$$

Since $i[V, \hat{S}^\alpha(\mathbf{r})] = 0$, $i[U, \hat{S}^\alpha(\mathbf{r})] = 0$, and $i[H_Z(\mathbf{Q}), \hat{S}^\alpha(\mathbf{r})] = 0$, we only need to consider $i[H_0, \hat{S}^\alpha(\mathbf{r})]$, $i[H_J, \hat{S}^\alpha(\mathbf{r})]$, and $i[H_{\mathrm{SO}}, \hat{S}^\alpha(\mathbf{r})]$. Let us first calculate $i[H_0, \hat{S}^\alpha(\mathbf{r})]$. By using the Fourier decomposition from Eq. (72) we obtain

$$i[H_0, \hat{S}^\alpha(\mathbf{r})] = i\frac{1}{\Omega} \sum_{\mathbf{q}} e^{i\mathbf{q}\cdot\mathbf{r}} \sum_i \frac{\hat{s}^\alpha_i}{2m} [\hat{\mathbf{p}}^2_i, e^{-i\mathbf{q}\cdot\hat{\mathbf{r}}_i}]. \tag{74}$$

From

$$[\hat{\mathbf{p}}_i, e^{-i\mathbf{q}\cdot\hat{\mathbf{r}}_i}] = -\mathbf{q} e^{-i\mathbf{q}\cdot\hat{\mathbf{r}}_i}, \tag{75}$$

we have

$$\begin{aligned}i[H_0, \hat{S}^\alpha(\mathbf{r})] &= i\frac{1}{\Omega} \sum_{\mathbf{q}} e^{i\mathbf{q}\cdot\mathbf{r}} \sum_i \frac{\hat{s}^\alpha_i}{2m} (-\mathbf{q}) \cdot \{\hat{\mathbf{p}}_i, e^{-i\mathbf{q}\cdot\hat{\mathbf{r}}_i}\} \\ &= -i\frac{1}{\Omega} \sum_{\mathbf{q}} e^{i\mathbf{q}\cdot\mathbf{r}} \mathbf{q} \cdot \mathbf{j}^\alpha_{\mathbf{q}} = -\nabla \cdot \mathbf{j}^\alpha(\mathbf{r}),\end{aligned} \tag{76}$$

where the spin current density operator is defined as $\mathbf{j}^\alpha(\mathbf{r}) = \sum_{i=1}^{N_e} \frac{1}{2m} \hat{s}^\alpha_i \{\hat{\mathbf{p}}_i, \delta(\hat{\mathbf{r}}_i - \mathbf{r})\}$ and its Fourier decomposition reads $\mathbf{j}^\lambda(\mathbf{r}) = \frac{1}{\Omega} \sum_{\mathbf{q}} e^{i\mathbf{q}\cdot\mathbf{r}} \mathbf{j}^\lambda_{\mathbf{q}}$, with $\mathbf{j}^\lambda_{\mathbf{q}} = \frac{1}{2m} \sum_i \hat{s}^\lambda_i \{\hat{\mathbf{p}}_i, e^{-i\mathbf{q}\cdot\hat{\mathbf{r}}_i}\}$ ($\lambda = x, y, z$). We note that above definition of spin currents emerges naturally in the present context, and must be carefully distinguished from the one commonly used in the literature [5–7], $\tilde{\mathbf{j}}^\alpha(\mathbf{r}) = \sum_{j=1}^{N_e} \frac{1}{2} \{\hat{\mathbf{v}}_j, \hat{s}^\alpha_j \delta(\hat{\mathbf{r}}_j - \mathbf{r})\}$, where the velocity operator, defined by $\hat{\mathbf{v}}_j = i[H, \hat{\mathbf{r}}_j]/\hbar$, picks up an additional anomalous spin-dependent term due to the presence of the SOI term $H_{\mathrm{SO}}$. This anomalous term is absent in $\mathbf{j}^\alpha(\mathbf{r})$.

Let us now calculate $i[H_J, \hat{S}^\alpha(\mathbf{r})]$,

$$\begin{aligned}i[H_J, \hat{S}^\alpha(\mathbf{r})] &= i[J \sum_{j,k} \hat{\mathbf{s}}_k \delta(\hat{\mathbf{r}}_k - \mathbf{R}_j) \hat{\mathbf{I}}_j, \sum_i \hat{s}^\alpha_i \delta(\hat{\mathbf{r}}_i - \mathbf{r})] = J \sum_{\beta=x,y,z} \sum_{i,j} i\delta(\hat{\mathbf{r}}_i - \mathbf{R}_j)\delta(\hat{\mathbf{r}}_i - \mathbf{r}) i\hat{I}^\beta_j \hat{s}^\gamma_i \epsilon_{\beta\alpha\gamma} \\ &= -J \sum_{\beta=x,y,z} \sum_j \delta(\mathbf{R}_j - \mathbf{r}) \hat{I}^\beta_j \hat{s}^\gamma_j \epsilon_{\beta\alpha\gamma} = J \sum_j \delta(\mathbf{R}_j - \mathbf{r})(\hat{\mathbf{I}}_j \wedge \hat{\mathbf{s}}_j)_\alpha.\end{aligned} \tag{77}$$

Finally we consider $i[H_{\mathrm{SO}}, \hat{S}^\alpha(\mathbf{r})]$. Since

$$i[H_R, \hat{S}^\alpha(\mathbf{r})] = i\frac{1}{\Omega} \sum_{\mathbf{q}} e^{i\mathbf{q}\cdot\mathbf{r}} [H_R, \hat{S}^\alpha_{\mathbf{q}}], \tag{78}$$

and

$$[H_R, \hat{S}^\alpha_{\mathbf{q}}] = \alpha \sum_i \underbrace{[\hat{p}^y_i \hat{s}^x_i, \sum_k e^{-i\mathbf{q}\cdot\hat{\mathbf{r}}_k} \hat{s}^\alpha_k]}_{=A} - \alpha \sum_i \underbrace{[\hat{p}^x_i \hat{s}^y_i, \sum_k e^{-i\mathbf{q}\cdot\hat{\mathbf{r}}_k} \hat{s}^\alpha_k]}_{=B}, \tag{79}$$

with

$$A = \hat{s}^x_i \hat{s}^\alpha_i [\hat{p}^y_i, e^{-i\mathbf{q}\cdot\hat{\mathbf{r}}_i}] + [\hat{s}^x_i, \hat{s}^\alpha_i] e^{-i\mathbf{q}\cdot\hat{\mathbf{r}}_i} \hat{p}^y_i = \frac{i}{2} \epsilon_{x\alpha\gamma} \hat{s}^\gamma_i \{\hat{p}^y_i, e^{-i\mathbf{q}\cdot\hat{\mathbf{r}}_i}\} \tag{80}$$

$$\Rightarrow \alpha \frac{i}{2} \epsilon_{x\alpha\gamma} \sum_i \hat{s}^\gamma_i \{\hat{p}^y_i, e^{-i\mathbf{q}\cdot\hat{\mathbf{r}}_i}\} = \frac{i}{2} \alpha \epsilon_{x\alpha\gamma} 2m j^\gamma_{\mathbf{q},y}, \tag{81}$$

$$B = \hat{s}_i^y \hat{s}_i^\alpha [\hat{p}_i^x, e^{-i\mathbf{q}\cdot\hat{\mathbf{r}}_i}] + [\hat{s}_i^y, \hat{s}_i^\alpha] e^{-i\mathbf{q}\cdot\hat{\mathbf{r}}_i} \hat{p}_i^x = \frac{i}{2}\epsilon_{y\alpha\gamma} \hat{s}_i^\gamma \{\hat{p}_i^x, e^{-i\mathbf{q}\cdot\hat{\mathbf{r}}_i}\} \tag{82}$$

$$\Rightarrow \frac{i}{2}\alpha\epsilon_{y\alpha\gamma} \sum_i \hat{s}_i^\gamma \{\hat{p}_i^x, e^{-i\mathbf{q}\cdot\hat{\mathbf{r}}_i}\} = \frac{i}{2}\alpha\epsilon_{y\alpha\gamma} 2mj_{\mathbf{q},x}^\gamma, \tag{83}$$

we have that

$$i[H_R, \hat{S}^\alpha(\mathbf{r})] = -\frac{1}{2}\alpha\epsilon_{x\alpha\gamma} 2m j_y^\gamma + \frac{1}{2}\alpha\epsilon_{y\alpha\gamma} 2m j_x^\gamma, \tag{84}$$

where we have used $j_\gamma^\lambda(\mathbf{r}) = \frac{1}{\Omega}\sum_{\mathbf{q}} e^{i\mathbf{q}\cdot\mathbf{r}} j_{\mathbf{q},\gamma}^\lambda$ ($\gamma = x,y,z$).

For the Dresselhaus term the calculation is similar. Since $i[H_D, \hat{S}^\alpha(\mathbf{r})] = i\frac{1}{\Omega}\sum_{\mathbf{q}} e^{i\mathbf{q}\cdot\mathbf{r}}[H_D, \hat{S}_{\mathbf{q}}^\alpha]$ and

$$[H_D, \hat{S}_{\mathbf{q}}^\alpha] = \beta \sum_i \underbrace{[\hat{p}_i^x \hat{s}_i^x, e^{-i\mathbf{q}\cdot\hat{\mathbf{r}}_i}\hat{s}_i^\alpha]}_{=A} - \beta \sum_i \underbrace{[\hat{p}_i^y \hat{s}_i^y, e^{-i\mathbf{q}\cdot\hat{\mathbf{r}}_i}\hat{s}_i^\alpha]}_{=B}, \tag{85}$$

with

$$A = \hat{s}_i^x \hat{s}_i^\alpha [\hat{p}_i^x, e^{-i\mathbf{q}\cdot\hat{\mathbf{r}}_i}] + [\hat{s}_i^x, \hat{s}_i^\alpha] e^{-i\mathbf{q}\cdot\hat{\mathbf{r}}_i} \hat{p}_i^x = \frac{i}{2}\epsilon_{x\alpha\gamma} \hat{s}_i^\gamma \{\hat{p}_i^x, e^{-i\mathbf{q}\cdot\hat{\mathbf{r}}_i}\} \tag{86}$$

$$\Rightarrow \frac{i}{2}\beta\epsilon_{x\alpha\gamma} \sum_i \hat{s}_i^\gamma \{\hat{p}_i^x, e^{-i\mathbf{q}\cdot\hat{\mathbf{r}}_i}\} = \frac{i}{2}\beta\epsilon_{x\alpha\gamma} 2mj_{\mathbf{q},x}^\gamma, \tag{87}$$

$$B = \hat{s}_i^y \hat{s}_i^\alpha [\hat{p}_i^y, e^{-i\mathbf{q}\cdot\hat{\mathbf{r}}_i}] + [\hat{s}_i^y, \hat{s}_i^\alpha] e^{-i\mathbf{q}\cdot\hat{\mathbf{r}}_i} \hat{p}_i^y = \frac{i}{2}\epsilon_{y\alpha\gamma} \hat{s}_i^\gamma \{\hat{p}_i^y, e^{-i\mathbf{q}\cdot\hat{\mathbf{r}}_i}\} \tag{88}$$

$$\Rightarrow \frac{i}{2}\beta\epsilon_{y\alpha\gamma} \sum_i \hat{s}_i^\gamma \{\hat{p}_i^y, e^{-i\mathbf{q}\cdot\hat{\mathbf{r}}_i}\} = \frac{i}{2}\beta\epsilon_{y\alpha\gamma} 2mj_{\mathbf{q},y}^\gamma, \tag{89}$$

we can conclude that

$$i[H_D, \hat{S}^\alpha(\mathbf{r})] = -\frac{1}{2}\beta\epsilon_{x\alpha\gamma} 2m j_x^\gamma + \frac{1}{2}\beta\epsilon_{y\alpha\gamma} 2m j_y^\gamma, \tag{90}$$

where we used $j_\gamma^\lambda(\mathbf{r}) = \frac{1}{\Omega}\sum_{\mathbf{q}} e^{i\mathbf{q}\cdot\mathbf{r}} j_{\mathbf{q},\gamma}^\lambda$ as above.

Let us now derive the Heisenberg equation of motion for the lattice spin density. Since $i[H_e, \hat{I}^\alpha(\mathbf{r})] = 0$, we only need to consider $i[H_J, \hat{I}^\alpha(\mathbf{r})]$ and $i[H_Z(\mathbf{Q}), \hat{I}^\alpha(\mathbf{r})]$. Let us first calculate $i[H_J, \hat{I}^\alpha(\mathbf{r})]$:

$$\begin{aligned}
i[H_J, \hat{I}^\alpha(\mathbf{r})] &= i[J \sum_{i,j} \hat{\mathbf{s}}_i \delta(\hat{\mathbf{r}}_i - \mathbf{R}_j)\hat{\mathbf{I}}_j, \sum_l \hat{I}_l^\alpha \delta(\mathbf{R}_l - \mathbf{r})] \\
&= iJ \sum_{\beta=x,y,z} \sum_{i,j} \delta(\hat{\mathbf{r}}_i - \mathbf{R}_j)\delta(\mathbf{R}_j - \mathbf{r}) i\hat{s}_i^\beta \hat{I}_j^\gamma \epsilon_{\beta\alpha\gamma} = -J \sum_{\beta=x,y,z} \sum_j \delta(\mathbf{R}_j - \mathbf{r}) \hat{s}_j^\beta \hat{I}_j^\gamma \epsilon_{\alpha\gamma\beta} \\
&= -J \sum_j \delta(\mathbf{R}_j - \mathbf{r})(\hat{\mathbf{I}}_j \times \hat{\mathbf{s}}_j)_\alpha = -i[H_J, \hat{S}^\alpha(\mathbf{r})].
\end{aligned} \tag{91}$$

For the (anti-)ferromagnetic case the Zeeman term is $H_Z(\mathbf{Q}) = h \sum_j \left( e^{-\mathbf{Q}\cdot\hat{\mathbf{R}}_j} \hat{I}_j^z + e^{i\mathbf{Q}\cdot\hat{\mathbf{R}}_j} \hat{I}_j^z \right)$, and

$$\begin{aligned}
i[H_Z(\mathbf{Q}), \hat{I}^\alpha(\mathbf{r})] &= i[h \sum_l \left( e^{-\mathbf{Q}\cdot\hat{\mathbf{R}}_l} \hat{I}_l^z + e^{i\mathbf{Q}\cdot\hat{\mathbf{R}}_l} \hat{I}_l^z \right), \sum_j \hat{I}_j^\alpha \delta(\mathbf{R}_j - \mathbf{r})] \\
&= ih \sum_{j,l} (e^{-i\mathbf{Q}\cdot\mathbf{R}_l} + e^{i\mathbf{Q}\cdot\mathbf{R}_l}) \delta(\mathbf{R}_j - \mathbf{r})[\hat{I}_l^z, \hat{I}_j^\alpha] \\
&= ih \sum_j (e^{-i\mathbf{Q}\cdot\mathbf{R}_j} + e^{i\mathbf{Q}\cdot\mathbf{R}_j}) \delta(\mathbf{R}_j - \mathbf{r}) i\epsilon_{z\alpha\gamma} \hat{I}_j^\gamma.
\end{aligned} \tag{92}$$

The continuity equation for the $z$-component of the total spin density takes the following form,

$$\dot{\Sigma}^z(\mathbf{r}) = -\nabla \cdot \mathbf{j}^z(\mathbf{r}) - \alpha\epsilon_{xzy} m j_y^y + \alpha\epsilon_{yzx} m j_x^x - \beta\epsilon_{xzy} m j_x^y + \beta\epsilon_{yzx} m j_y^x, \tag{93}$$

$$\Rightarrow \dot{\Sigma}^z(\mathbf{r}) + \nabla \cdot \mathbf{j}^z(\mathbf{r}) = m\alpha(j_y^y + j_x^x) + m\beta(j_x^y + j_y^x). \tag{94}$$

In the homogeneous and stationary limit, the left-hand side of Eq. (93) vanishes and this leads to

$$\langle j_x^x \rangle = -\langle j_y^y \rangle \quad \text{and} \quad \langle j_x^y \rangle = -\langle j_y^x \rangle. \tag{95}$$

For the helical case, the Zeeman term is $\widetilde{H}_Z(\mathbf{Q}) = \sqrt{2/3}h \sum_j \left( e^{-i\mathbf{Q}\cdot\mathbf{R}_j} \hat{I}_j^+ + e^{i\mathbf{Q}\cdot\mathbf{R}_j} \hat{I}_j^- \right)$ and

$$\begin{aligned} i[\widetilde{H}_Z(\mathbf{Q}), \hat{I}^\alpha(\mathbf{r})] &= i[\sqrt{2/3}h \sum_l \left( e^{-i\mathbf{Q}\cdot\mathbf{R}_l} \hat{I}_l^+ + e^{i\mathbf{Q}\cdot\mathbf{R}_l} \hat{I}_l^- \right), \sum_j \hat{I}_j^\alpha \delta(\mathbf{R}_j - \mathbf{r})] \\ &= i\sqrt{2/3}h \sum_j \left( e^{-i\mathbf{Q}\cdot\mathbf{R}_j} \delta(\mathbf{R}_j - \mathbf{r})[\hat{I}_j^+, \hat{I}_j^\alpha] + e^{i\mathbf{Q}\cdot\mathbf{R}_j} \delta(\mathbf{R}_j - \mathbf{r})[\hat{I}_j^-, \hat{I}_j^\alpha] \right) \\ &= i\sqrt{2/3}h \sum_j e^{-i\mathbf{Q}\cdot\mathbf{R}_j} \delta(\mathbf{R}_j - \mathbf{r})(i\epsilon_{x\alpha\gamma}\hat{I}_j^\gamma - \epsilon_{y\alpha\gamma}\hat{I}_j^\gamma) + i\sqrt{2/3}h \sum_j e^{i\mathbf{Q}\cdot\mathbf{R}_j} \delta(\mathbf{R}_j - \mathbf{r})(i\epsilon_{x\alpha\gamma}\hat{I}_j^\gamma + \epsilon_{y\alpha\gamma}\hat{I}_j^\gamma). \end{aligned}$$
(96)

For $\alpha = z$ we thus have

$$i[\widetilde{H}_Z(\mathbf{Q}), \hat{I}^\alpha(\mathbf{r})] = i\sqrt{2/3}h \sum_j \left( e^{-i\mathbf{Q}\cdot\mathbf{R}_j} \delta(\mathbf{R}_j - \mathbf{r})(-\hat{I}_j^+) + e^{i\mathbf{Q}\cdot\mathbf{R}_j} \delta(\mathbf{R}_j - \mathbf{r})\hat{I}_j^- \right). \tag{97}$$

In this case the continuity equation for the $z$-component of the total spin density reads

$$\dot{\Sigma}^z(\mathbf{r}) + \nabla \cdot \mathbf{j}^z(\mathbf{r}) = m\alpha(j_y^y + j_x^x) + m\beta(j_x^y + j_y^x) + i\sqrt{2/3}h \sum_j \left( e^{-i\mathbf{Q}\cdot\mathbf{R}_j} \delta(\mathbf{R}_j - \mathbf{r})(-\hat{I}_j^+) + e^{i\mathbf{Q}\cdot\mathbf{R}_j} \delta(\mathbf{R}_j - \mathbf{r})\hat{I}_j^- \right). \tag{98}$$

In the presence of the lattice spin-spin interaction term $H_\mathcal{I}$, the right-hand side of Eqs. (93) and (98) acquires an additional term

$$\begin{aligned} i[H_\mathcal{I}, \hat{I}^\alpha(\mathbf{r})] &= i[\sum_{j,l} \sum_{\beta=x,y,z} \mathcal{I}_{jl} \hat{I}_j^\beta \hat{I}_l^\beta, \sum_k \hat{I}_k^\alpha \delta(\mathbf{R}_k - \mathbf{r})] = i\sum_{j,l,k,\beta} \mathcal{I}_{jl} \delta(\mathbf{R}_k - \mathbf{r}) \left( \hat{I}_j^\beta [\hat{I}_l^\beta, \hat{I}_k^\alpha] + [\hat{I}_j^\beta, \hat{I}_k^\alpha] \hat{I}_l^\beta \right) \\ &= i\sum_{j,l,\beta} \mathcal{I}_{jl} \left( \delta(\mathbf{R}_l - \mathbf{r}) \hat{I}_j^\beta i\epsilon_{\beta\alpha\gamma} \hat{I}_l^\gamma + \delta(\mathbf{R}_j - \mathbf{r}) i\epsilon_{\beta\alpha\gamma} \hat{I}_j^\gamma \hat{I}_l^\beta \right) \\ &= \sum_{j,l,\beta} \mathcal{I}_{jl} \delta(\mathbf{R}_l - \mathbf{r}) \epsilon_{\alpha\beta\gamma} \hat{I}_j^\beta \hat{I}_l^\gamma + \sum_{j,l,\beta} \mathcal{I}_{jl} \delta(\mathbf{R}_j - \mathbf{r}) \epsilon_{\alpha\beta\gamma} \hat{I}_l^\beta \hat{I}_j^\gamma = 2\sum_{jl} \mathcal{I}_{jl} \delta(\mathbf{R}_l - \mathbf{r}) (\hat{I}_j \times \hat{I}_l)_\alpha. \end{aligned} \tag{99}$$

## VII. EQUILIBRIUM SPIN-CURRENTS FOR $U, V, J, T = 0$

The aim of this section is to calculate uniform equilibrium spin currents for a system described by Hamiltonian (1) in the special case $V, U, J, T = 0$. The calculation follows the ones given in Refs. [5, 7]; however, due to the spin currents occcurring here without anomalous velocity term (see text after Eq. (76)), the leading term will turn out to be first order in the SOIs (instead of 3rd order [5, 7]). The single-particle Hamiltonian we consider takes the following form,

$$H = \frac{\hat{\mathbf{p}}^2}{2m} + \frac{\alpha}{\hbar}(\sigma_x \hat{p}_y - \sigma_y \hat{p}_x) + \frac{\beta}{\hbar}(\sigma_y \hat{p}_y + \sigma_x \hat{p}_x), \tag{100}$$

where $\hat{\mathbf{p}}$ is the electron momentum operator and $\sigma_{x,y,z}$ are the usual Pauli matrices. The eigenstates of the system are

$$\psi_{\mathbf{k},s}(\mathbf{r}) = \frac{e^{i\mathbf{k}\cdot\mathbf{r}}}{\sqrt{A}} \mathbf{u}_s(\mathbf{k}), \tag{101}$$

where $A$ is the area of the system and $\mathbf{u}_s(\mathbf{k}) = \begin{pmatrix} 1 \\ s\frac{\Gamma^+}{|\Gamma^+|} \end{pmatrix}$, with $\mathbf{\Gamma} = \frac{1}{\sqrt{2}} \begin{pmatrix} -\beta k_x + \alpha k_y \\ \beta k_y - \alpha k_x \\ 0 \end{pmatrix}$. The spectrum is given by

$$E_s(k_x, k_y) = \frac{\hbar^2 k^2}{2m} + s\sqrt{(\alpha^2 + \beta^2)k^2 - 4k_x k_y \alpha\beta}, \tag{102}$$

and it takes the following form in polar coordinates $(k_x, k_y) = (k\cos(\theta), k\sin(\theta))$:

$$E_s(k,\theta) = \frac{\hbar^2 k^2}{2m} + sk\underbrace{\sqrt{(\alpha^2+\beta^2) - 4\cos(\theta)\sin(\theta)\alpha\beta}}_{=:\alpha(\theta)}. \tag{103}$$

The Fermi wavevectors $k_\pm(\theta)$ of the two branches are defined by $E_F = E_{s=\mp 1}(k_\pm(\theta)) = \frac{\hbar^2 k_\pm(\theta)^2}{2m} + sk_\pm(\theta)\alpha(\theta)$ and the explicit expressions read

$$k_\pm(\theta) = \pm\frac{\alpha(\theta)m}{\hbar^2} + \sqrt{\left(\frac{m}{\hbar^2}\alpha(\theta)\right)^2 + \frac{2m}{\hbar^2}E_F}. \tag{104}$$

In the perturbative limit where $m\alpha(\theta)^2 \ll \hbar^2 E_F$, we expand $k_\pm(\theta)$ up to second order in $\alpha(\theta)$ and obtain

$$k_\pm(\theta) = \sqrt{\frac{2m}{\hbar^2}E_F}\frac{m}{4\hbar^2 E_F}\alpha(\theta)^2 \pm \frac{m}{\hbar^2}\alpha(\theta) + \sqrt{\frac{2m}{\hbar^2}E_F}. \tag{105}$$

The spin-current tensor $\mathcal{T}_{lm}$ is given by

$$\mathcal{T}_{lm} = \frac{1}{\Omega}\langle j^m_{\mathbf{q}=\mathbf{0},l}\rangle_0 = \frac{1}{2}\sum_s\int\frac{d^2k}{(2\pi)^2}\langle \sigma_m v_l + v_l\sigma_m\rangle_{\mathbf{k},s}, \tag{106}$$

where $\langle...\rangle_{\mathbf{k},s}$ denotes the expectation value in the eigenstates $\psi_{\mathbf{k},s}$ and the integration must be performed over $k \leq k_\pm(\theta)$.

Let us first calculate $\mathcal{T}_{xy}$. Since $v_x = \hbar k_x/m$, we have that $\{\sigma_y, v_x\} = 2\hbar k_x\sigma_y/m$, and

$$\frac{2\hbar}{m}\langle\sigma_y k_x\rangle_{\mathbf{k},s} = \frac{2\hbar}{m}s\frac{k_x\Gamma_y}{|\Gamma^+|} = \frac{2\hbar}{m}s\frac{k_x(\beta k_y - \alpha k_x)}{\sqrt{k^2(\alpha^2+\beta^2) - 4k_xk_y\alpha\beta}}. \tag{107}$$

With the use of polar coordinates we obtain

$$\frac{2\hbar}{m}\langle\sigma_y k_x\rangle_{\mathbf{k},s} = \frac{2\hbar}{m}s\frac{\beta k^2\cos(\theta)\sin(\theta) - \alpha k^2\cos^2(\theta)}{k\underbrace{\sqrt{(\alpha^2+\beta^2) - 4\cos(\theta)\sin(\theta)\alpha\beta}}_{=\alpha(\theta)}}. \tag{108}$$

From Eqs. (106,108) we obtain

$$\begin{aligned}\mathcal{T}_{xy} &= \frac{1}{2}\frac{2\hbar}{m}\frac{1}{(2\pi)^2}\int_{k_+(\theta)}^{k_-(\theta)}\int_0^{2\pi} dkd\theta k\frac{\beta k\cos(\theta)\sin(\theta) - \alpha k\cos^2(\theta)}{\alpha(\theta)}\\ &= \frac{\hbar}{m}\frac{1}{(2\pi)^2}\int_0^{2\pi}d\theta\frac{k_-(\theta)^3 - k_+(\theta)^3}{3}\frac{\beta\cos(\theta)\sin(\theta) - \alpha\cos^2(\theta)}{\alpha(\theta)}\\ &= -\frac{\hbar}{m}\frac{1}{(2\pi)^2}\frac{1}{3}\int_0^{2\pi}d\theta[\frac{8m^3}{\hbar^6}\alpha(\theta)^3 + \frac{12m^2}{\hbar^4}E_F\alpha(\theta)]\frac{\beta\cos(\theta)\sin(\theta) - \alpha\cos^2(\theta)}{\alpha(\theta)}.\end{aligned} \tag{109}$$

Let us now split the above integral in two parts:

$$\begin{aligned}I_1 &= -\frac{\hbar}{m}\frac{1}{(2\pi)^2}\frac{1}{3}\frac{8m^3}{\hbar^6}\int_0^{2\pi}d\theta\alpha(\theta)^2(\beta\cos(\theta)\sin(\theta) - \alpha\cos^2(\theta))\\ &= -\frac{2m^2}{3\pi^2\hbar^5}\int_0^{2\pi}d\theta(\alpha^2+\beta^2 - 4\cos(\theta)\sin(\theta)\alpha\beta)(\beta\cos(\theta)\sin(\theta) - \alpha\cos^2(\theta))\\ &= \frac{2m^2}{3\pi\hbar^5}\alpha(\alpha^2+2\beta^2),\end{aligned} \tag{110}$$

$$\begin{aligned}I_2 &= -\frac{\hbar}{m}\frac{1}{(2\pi)^2}\frac{1}{3}\frac{12m^2}{\hbar^4}E_F\int_0^{2\pi}d\theta(\beta\cos(\theta)\sin(\theta) - \alpha\cos^2(\theta))\\ &= \frac{mE_F}{\pi\hbar^3}\alpha.\end{aligned} \tag{111}$$

The spin-current $\mathcal{T}_{xy}$ is thus given by

$$\mathcal{T}_{xy} \approx \frac{2m^2}{3\pi\hbar^5}\alpha(\alpha^2 + 2\beta^2) + \frac{mE_F}{\pi\hbar^3}\alpha. \tag{112}$$

Note that the term cubic in SOI agrees with earlier results [5, 7], while the linear term survives here due to the absence of an anomalous velocity term. However, since $m\alpha(\theta)^2 \ll \hbar^2 E_F$ the cubic term is negligible and we obtain

$$\mathcal{T}_{xy} \approx \frac{mE_F}{\pi\hbar^3}\alpha. \tag{113}$$

A similar calculation shows that

$$\mathcal{T}_{yy} \approx -\frac{mE_F}{\pi\hbar^3}\beta. \tag{114}$$

Note that differently from the main text, both Hamiltonian (100) and the spin current density are expressed in terms of Pauli matrices and not in terms of spin operators. Therefore, we need to multiply our results by a factor $(\hbar/2)^2$. Another multiplication with a factor $\hbar$ arises from the fact that the spin-orbit part is multiplied by $1/\hbar$ in (100) as compared to our definition in the main text,

$$\mathcal{T}_{xy} \approx \frac{mE_F}{4\pi}\alpha, \tag{115}$$

$$\mathcal{T}_{yy} \approx -\frac{mE_F}{4\pi}\beta. \tag{116}$$

## VIII. SPIN-ORBIT INTERACTION WITH $\alpha = \beta$

Let us consider the special case when Rashba and Dresselhaus coefficients are equal, i.e., $\alpha = \beta$. In such a case, the spin-orbit Hamiltonian takes the following form

$$H_{\text{SO}} = \alpha \sum_i (\hat{p}_i^x + \hat{p}_i^y)(\hat{s}_i^x - \hat{s}_i^y), \tag{117}$$

and we define a gauge transformation $U = e^{i\sum_k \hat{\mathbf{A}}_k \cdot \hat{\mathbf{r}}_k}$ with gauge vector field $\mathbf{A}_k = (-\alpha m(\hat{s}_k^x - \hat{s}_k^y), -\alpha m(\hat{s}_k^x - \hat{s}_k^y), 0)$. Since

$$U\hat{p}_i^{x,y}U^{-1} = e^{i\sum_k \hat{\mathbf{A}}_k \cdot \hat{\mathbf{r}}_k}\hat{p}_i^{x,y}e^{-i\sum_k \hat{\mathbf{A}}_k \cdot \hat{\mathbf{r}}_k} = \hat{p}_i^{x,y} + \alpha m(\hat{s}_i^x - \hat{s}_i^y), \tag{118}$$

we have that

$$U\sum_i \frac{\hat{\mathbf{p}}_i^2}{2m}U^{-1} = \sum_i \frac{(\hat{p}_i^x + \alpha m(\hat{s}_i^x - \hat{s}_i^y))^2 + (\hat{p}_i^y + \alpha m(\hat{s}_i^x - \hat{s}_i^y))^2}{2m} = \sum_i \frac{\hat{\mathbf{p}}_i^2}{2m} + H_{\text{SO}} + \frac{\alpha^2 m}{2}, \tag{119}$$

where the constant $\alpha^2 m/2$ can be neglected without loss of generality. Under gauge transformation the spin isotropic term $H_J = J\sum_{j=1}^{N_I} \hat{\mathbf{S}}_j \cdot \hat{\mathbf{I}}_j$ transforms as follows,

$$\begin{aligned}
UH_J U^{-1} &= UJ\sum_{j=1}^{N_I} \hat{\mathbf{S}}_j \cdot \hat{\mathbf{I}}_j U^{-1} = J\sum_{j=1}^{N_I} U\hat{\mathbf{S}}_j U^{-1} \cdot \hat{\mathbf{I}}_j = J\sum_{j=1}^{N_I} e^{i\sum_k \hat{\mathbf{A}}_k \cdot \hat{\mathbf{r}}_k}\sum_{i=1}^{N_e} \hat{\mathbf{s}}_i \delta(\hat{\mathbf{r}}_i - \mathbf{R}_j)e^{-i\sum_k \hat{\mathbf{A}}_k \cdot \hat{\mathbf{r}}_k}\hat{\mathbf{I}}_j \\
&= J\sum_{j=1}^{N_I}\sum_{i=1}^{N_e} \underbrace{e^{i\sum_k \hat{\mathbf{A}}_k \cdot \hat{\mathbf{r}}_k}\hat{\mathbf{s}}_i e^{-i\sum_k \hat{\mathbf{A}}_k \cdot \hat{\mathbf{r}}_k}}_{=\mathcal{R}(\hat{\mathbf{r}}_i)\hat{\mathbf{s}}_i}\delta(\hat{\mathbf{r}}_i - \mathbf{R}_j)\hat{\mathbf{I}}_j = J\sum_{j=1}^{N_I}\sum_{i=1}^{N_e} \delta(\hat{\mathbf{r}}_i - \mathbf{R}_j)\mathcal{R}(\mathbf{R}_j)\hat{\mathbf{s}}_i \cdot \hat{\mathbf{I}}_j \\
&= J\sum_{j=1}^{N_I}\sum_{i=1}^{N_e} \delta(\hat{\mathbf{r}}_i - \mathbf{R}_j)\hat{\mathbf{s}}_i \cdot \underbrace{\mathcal{R}(\mathbf{R}_j)^T \hat{\mathbf{I}}_j}_{=:\hat{\mathbf{I}}_j^{\mathcal{R}^T}} = J\sum_{j=1}^{N_I}\sum_{i=1}^{N_e} \delta(\hat{\mathbf{r}}_i - \mathbf{R}_j)\hat{\mathbf{s}}_i \cdot \hat{\mathbf{I}}_j^{\mathcal{R}^T},
\end{aligned} \tag{120}$$

where $\mathcal{R} \in SO(3)$ is a $3 \times 3$ special orthogonal matrix. The lattice spin Hamiltonian $H_\mathcal{I}$ remains unchanged under gauge transformation. From this, we can conclude that Hamiltonian $H = H_e + H_\mathcal{I} + H_J + H_{\text{SO}}$ and Hamiltonian

$$H^{\mathcal{R}^T} = \underbrace{\sum_i \frac{\hat{\mathbf{p}}_i^2}{2m} + \sum_{i<j} V(\hat{\mathbf{r}}_i - \hat{\mathbf{r}}_j) + \sum_i U(\hat{\mathbf{r}}_i)}_{=H_e} + \underbrace{\sum_{j,l} \mathcal{I}_{jl} \hat{\mathbf{I}}_j \cdot \hat{\mathbf{I}}_l}_{=H_\mathcal{I}} + J \sum_{j=1}^{N_I} \hat{\mathbf{S}}_j \cdot \hat{\mathbf{I}}_j^{\mathcal{R}^T}, \tag{121}$$

are equivalent, i.e., $U^{-1} H^{\mathcal{R}^T} U = H$. Let us now introduce an impurity spin rotation $\widetilde{U}$ such that $\widetilde{U}^{-1} \hat{\mathbf{I}}_j \widetilde{U} = \hat{\mathbf{I}}_j^{\mathcal{R}^T}$:

$$\widetilde{U}^{-1} H_0 \widetilde{U} = H^{\mathcal{R}^T}, \tag{122}$$

where $H_0 = H_e + H_\mathcal{I} + \sum_{j=1}^{N_I} \hat{\mathbf{S}}_j \cdot \hat{\mathbf{I}}_j$. We conclude that

$$U^{-1} \widetilde{U}^{-1} H_0 \widetilde{U} U = H. \tag{123}$$

The canonical ensemble average of an operator $\mathcal{O}$ is given by $\langle \mathcal{O} \rangle_H = \text{Tr}(e^{-\beta H} \mathcal{O})/\text{Tr}(e^{-\beta H})$. Therefore, if we use the fact that $U^{-1} \widetilde{U}^{-1} H_0 \widetilde{U} U = H$, we obtain

$$\langle \mathcal{O} \rangle_H = \text{Tr}\left(e^{-\beta \widetilde{U} U H U^{-1} \widetilde{U}^{-1}} \left[\widetilde{U} U \mathcal{O} U^{-1} \widetilde{U}^{-1}\right]\right)/\text{Tr}\left(e^{-\beta \widetilde{U} U H U^{-1} \widetilde{U}^{-1}}\right) = \text{Tr}\left(e^{-\beta H_0} \left[\widetilde{U} U \mathcal{O} U^{-1} \widetilde{U}^{-1}\right]\right)/\text{Tr}\left(e^{-\beta H_0}\right)$$
$$= \langle \widetilde{U} U \mathcal{O} U^{-1} \widetilde{U}^{-1} \rangle_{H_0}. \tag{124}$$

For the impurity spin magnetization $\hat{m}_I$ Eq. (124) implies that

$$\langle \hat{m}_I \rangle_H = \text{Tr}\left(e^{-\beta H_0} \hat{m}_I^\mathcal{R}\right)/\text{Tr}\left(e^{-\beta H_0}\right) = \langle \hat{m}_I^\mathcal{R} \rangle_{H_0}, \tag{125}$$

where $\hat{m}_I^\mathcal{R} = \widetilde{U} U \hat{m}_I U^{-1} \widetilde{U}^{-1} = \widetilde{U} \hat{m}_I \widetilde{U}^{-1}$ is defined analogously as $\hat{m}_I$ but with rotated spins $\hat{\mathbf{I}}_j^\mathcal{R} = \mathcal{R}(\mathbf{R}_j) \hat{\mathbf{I}}_j$. The explicit form of the matrix $\mathcal{R}(\mathbf{R}_j)$ is given by

$$\mathcal{R}(\mathbf{R}_j) = \begin{pmatrix} 1/2 + 1/2 \cos(\theta_j) & -1/2 + 1/2 \cos(\theta_j) & -1/\sqrt{2} \sin(\theta_j) \\ -1/2 + 1/2 \cos(\theta_j) & 1/2 + 1/2 \cos(\theta_j) & -1/\sqrt{2} \sin(\theta_j) \\ 1/\sqrt{2} \sin(\theta_j) & 1/\sqrt{2} \sin(\theta_j) & \cos(\theta_j) \end{pmatrix}, \tag{126}$$

and it corresponds to a rotation around the axis $\mathbf{n} = (1/\sqrt{2}, -1/\sqrt{2}, 0)$ by an angle of $\theta_j = -\sqrt{2} \alpha m (R_j^x + R_j^y)$. Note that the lattice vectors $\mathbf{R}_j = (R_j^x, R_j^y, R_j^z)$ should not be confused with the rotation matrix $\mathcal{R}$.

### A. (Anti-) ferromagnetic ordering

The aim of this section is to rule out ferromagnetic ordering in a system defined by Hamiltonian $H = H_e + H_\mathcal{I} + H_J + H_{\text{SO}} + H_Z(\mathbf{Q})$, where $H_{\text{SO}}$ is defined as in Eq. (117) and the Zeeman term is chosen to be $H_Z(\mathbf{Q}) = h \sum_{j=1}^{N_I} (e^{-i\mathbf{Q} \cdot \mathbf{R}_j} \hat{I}_j^z + e^{i\mathbf{Q} \cdot \mathbf{R}_j} \hat{I}_j^z)$. The magnetization is given by $m_I^z(\mathbf{Q}) = \langle \hat{m}_I^z(\mathbf{Q}) \rangle_H = \frac{1}{N_I} \langle \sum_{j=1}^{N_I} (e^{-i\mathbf{Q} \cdot \mathbf{R}_j} \hat{I}_j^z + e^{i\mathbf{Q} \cdot \mathbf{R}_j} \hat{I}_j^z) \rangle_H$. We know that $H_Z^\mathcal{R}(\mathbf{Q}) = \widetilde{U} U H_Z(\mathbf{Q}) U^{-1} \widetilde{U}^{-1} = h \sum_{j=1}^{N_I} (e^{-i\mathbf{Q} \cdot \mathbf{R}_j} (\hat{I}_j^\mathcal{R})^z + e^{i\mathbf{Q} \cdot \mathbf{R}_j} (\hat{I}_j^\mathcal{R})^z)$ and from Eq. (125) that $\langle \hat{m}_I^z(\mathbf{Q}) \rangle_H = \langle (\hat{m}_I^\mathcal{R})^z(\mathbf{Q}) \rangle_{H_0} = \frac{1}{N_I} \langle \sum_{j=1}^{N_I} (e^{-i\mathbf{Q} \cdot \mathbf{R}_j} (\hat{I}_j^\mathcal{R})^z + e^{i\mathbf{Q} \cdot \mathbf{R}_j} (\hat{I}_j^\mathcal{R})^z) \rangle_{H_0}$, with $H_0 = H_e + H_\mathcal{I} + H_J + H_Z^\mathcal{R}(\mathbf{Q})$. Since the rotation matrix $\mathcal{R}$ corresponds to a rotation around axis $\mathbf{n} = (1/\sqrt{2}, -1/\sqrt{2}, 0)$ (see Eq. (126)), we define a new coordinate system, namely: $\mathbf{z} \to \mathbf{z}' = (1/\sqrt{2}, -1/\sqrt{2}, 0)$, $\mathbf{y} \to \mathbf{y}' = (1/\sqrt{2}, 1/\sqrt{2}, 0)$, and $\mathbf{x} \to \mathbf{x}' = \mathbf{z} = (0,0,1)$. In this new coordinate system we obtain

$$H_Z^\mathcal{R}(\mathbf{Q}) = \frac{h}{2} \sum_{j=1}^{N_I} \left(e^{-i(\mathbf{Q}'+\mathbf{Q}) \cdot \mathbf{R}_j} \hat{I}_j^{+'} + e^{i(\mathbf{Q}'-\mathbf{Q}) \cdot \mathbf{R}_j} \hat{I}_j^{-'} + e^{i(\mathbf{Q}-\mathbf{Q}') \cdot \mathbf{R}_j} \hat{I}_j^{+'} + e^{i(\mathbf{Q}'+\mathbf{Q}) \cdot \mathbf{R}_j} \hat{I}_j^{-'}\right), \tag{127}$$

$$\langle (m_I^\mathcal{R})^z(\mathbf{Q}) \rangle_{H_0} = \frac{1}{2N_I} \left\langle \sum_{j=1}^{N_I} \left(e^{-i(\mathbf{Q}'+\mathbf{Q}) \cdot \mathbf{R}_j} \hat{I}_j^{+'} + e^{i(\mathbf{Q}'-\mathbf{Q}) \cdot \mathbf{R}_j} \hat{I}_j^{-'} + e^{i(\mathbf{Q}-\mathbf{Q}') \cdot \mathbf{R}_j} \hat{I}_j^{+'} + e^{i(\mathbf{Q}'+\mathbf{Q}) \cdot \mathbf{R}_j} \hat{I}_j^{-'}\right) \right\rangle_{H_0}, \tag{128}$$

where $\hat{I}_j^{\pm'} = \hat{I}_j^{x'} \pm i \hat{I}_j^{y'}$ and $\mathbf{Q}' = (-\sqrt{2} \alpha m, -\sqrt{2} \alpha m, 0)$. Note that in the new coordinate system $x'y'z'$, $H_\mathcal{I}$ and $H_J$ does not change their form.

The proof in section IV can be generalized in a straightforward way to a slightly different choice of operator $\widetilde{A}_{\mathbf{q}} = \hat{I}^+_{\mathbf{q}+\mathbf{K}} - \hat{I}^-_{\mathbf{q}-\mathbf{K}} + \hat{I}^+_{\mathbf{q}+\mathbf{K}'} - \hat{I}^-_{\mathbf{q}-\mathbf{K}'}$ and symmetry-breaking Zeeman term $H_Z(\mathbf{K}, \mathbf{K}') = h \sum_{j=1}^{N_I} (e^{-i\mathbf{K}\cdot\mathbf{R}_j} \hat{I}_j^{+'} + e^{i\mathbf{K}\cdot\mathbf{R}_j} \hat{I}_j^{-'} + e^{-i\mathbf{K}'\cdot\mathbf{R}_j} \hat{I}_j^{+'} + e^{i\mathbf{K}'\cdot\mathbf{R}_j} \hat{I}_j^{-'})$, with $\mathbf{K} = \mathbf{Q} + \mathbf{Q}'$ and $\mathbf{K}' = \mathbf{Q}' - \mathbf{Q}$. We thus conclude that in the limit of vanishing external magnetic field ($h \to 0$) $\langle (\hat{m}_I^{\mathcal{R}})^z \rangle_{H_0} = 0$ and consequently that $\langle \hat{m}_I^z \rangle_H = 0$. Since the proof is valid for any possible $\mathbf{Q}$, we have ruled out (anti-) ferromagnetic ordering along the $z$-direction in this case, too.

### B. Helical Ordering

Here we rigorously rule out helical ordering in the direction of $\mathbf{n} = (1/\sqrt{2}, -1/\sqrt{2}, 0)$ for a system described by the Hamiltonian $H = H_e + H_\mathcal{I} + H_J + H_{SO} + H_Z^{\mathcal{R}^T}$, where $H_{SO}$ is defined as in Eq. (117) and the Zeeman term is $H_Z^{\mathcal{R}^T} = h \sum_{j=1}^{N_I} (\hat{I}_j^{\mathcal{R}^T})^z$. The magnetization is given by $(m_I^{\mathcal{R}^T})^z = \langle (\hat{m}_I^{\mathcal{R}^T})^z \rangle_H = \frac{1}{N_I} \langle \sum_{j=1}^{N_I} (\hat{I}_j^{\mathcal{R}^T})^z \rangle_H$. From $\widetilde{U} U H_Z^{\mathcal{R}^T} U^{-1} \widetilde{U}^{-1} = H_Z = h \sum_{j=1}^{N_I} \hat{I}_j^z$ and Eq. (125), we have

$$\langle (\hat{m}_I^{\mathcal{R}^T})^z \rangle_H = \langle \hat{m}_I^z \rangle_{H_0}, \tag{129}$$

where $H_0 = H_e + H_\mathcal{I} + H_J + H_Z$. From the proof in section III, we know that in in the limit of vanishing external magnetic field ($h \to 0$) $\langle \hat{m}_I^z \rangle_{H_0} = 0$. From Eq. (129) we can thus conclude that $\langle (\hat{m}_I^{\mathcal{R}^T})^z \rangle_H = 0$ for $h \to 0$. In order to have a better idea of the type of helical order excluded here, we can rewrite the magnetization $(m_I^{\mathcal{R}^T})^z$ in the transformed coordinate system $x'y'z'$ defined in the paragraph above,

$$(m_I^{\mathcal{R}^T})^z = \frac{1}{2N_I} \left\langle \sum_{j=1}^{N_I} \left( e^{-i\mathbf{Q}\cdot\mathbf{R}_j} \hat{I}_j^{+'} + e^{i\mathbf{Q}\cdot\mathbf{R}_j} \hat{I}_j^{-'} \right) \right\rangle_H, \tag{130}$$

where $\mathbf{Q} = (\sqrt{2}\alpha m, \sqrt{2}\alpha m, 0)$. This corresponds to a helix in the $x'y'$-plane.

---